\documentclass[journal]{IEEEtran}
\usepackage{stfloats}
\newcommand{\ls}[1] {\dimen0=\fontdimen6\the\font \lineskip=#1\dimen0
\advance\lineskip.5\fontdimen5\the\font \advance\lineskip-\dimen0
\lineskiplimit=.9\lineskip \baselineskip=\lineskip
\advance\baselineskip\dimen0 \normallineskip\lineskip
\normallineskiplimit\lineskiplimit \normalbaselineskip\baselineskip
\ignorespaces } \ls{1}
\usepackage{color,graphicx}

\usepackage{times,float}

\usepackage{amsmath,amssymb}
\usepackage{subfigure,epsfig}
\usepackage{cite}
\newcommand {\beq} {\begin{equation}}
\newcommand {\eeq} {\end{equation}}
\newcommand {\barr} {\begin{array}}
\newcommand {\earr} {\end{array}}
\newcommand {\bear} {\begin{eqnarray}}
\newcommand {\eear} {\end{eqnarray}}
\newcommand {\bears} {\begin{eqnarray*}}
\newcommand {\eears} {\end{eqnarray*}}
\def \E{{\mathbb E}}
\def \P{{\mathbb P}}

\def\N{{\mathbb N}}

\def\bone{{\mathrm 1\!\!I}}
\newcommand{\eat}[1]{}
\def\bone{{\textbf 1}}
\newtheorem{remark}{Remark}
\begin{document}

\title{ Sleep Mode Analysis via Workload Decomposition 
}
\author{Amar Prakash Azad 
\thanks{ 
e-mail: amar.azad@ieee.org}
}
\maketitle
\begin{abstract}
The goal of this paper is to establish a general approach for analyzing queueing models with repeated inhomogeneous vacations. The server goes on for a vacation if the inactivity prolongs more than the vacation trigger duration. Once the system enters in vacation mode, it may continue for several consecutive vacations. At the end of a vacation, the server goes on another vacation, possibly with a {\em different} probability distribution; if during the previous vacation there have been no arrivals. However the system enters in vacation mode only if the inactivity is persisted beyond defined trigger duration. In order to get an insight on the influence of parameters on the performance, we choose to study a simple $M/G/1$ queue
(Poisson arrivals and general independent service times) which has the advantage of being tractable analytically. The theoretical model is applied to the problem of power saving for mobile devices in which the sleep durations of a device correspond to the vacations of the server. Various system performance metrics such as the frame response time and the economy of energy are derived. A constrained optimization problem is formulated to maximize the economy of energy achieved in power save mode, with constraints as QoS conditions to be met. An illustration of the proposed methods is shown with a WiMAX system scenario to obtain design parameters for better performance. Our analysis allows us not only to optimize the system parameters for a given traffic intensity but also to propose parameters that provide the best performance under worst case conditions.
\end{abstract}
\begin{IEEEkeywords}
M/G/1 queue with repeated vacations, power save mode, system
response time, gain optimization.
\end{IEEEkeywords}
\section{Introduction}
\label{s:intro}
Power save/sleep mode operation is the key point for energy efficient usage of mobile devices driven by limited battery lifetime. Current standards of Mobile communication such as WiFi, 3G and WiMAX have provisions to operate the mobile station in power save mode in case of low uses scenarios. A mobile operating in power save or sleep mode saves the battery energy and enhances lifetime but it also introduces unwanted delay in serving data packets arriving during sleep duration. Though energy is a major aspect for handheld devices, delays may also be crucial for various QoS services such as voice and video traffic. Mobility extension of WiMAX ~ \cite{man} is one of the most recent technologies whose sleep mode operation is being discussed in detail and standardized. 

The IEEE $802.16e$ standard ~\cite{man} defines 3 types of power saving classes.
\begin{itemize}
\item
Type I classes are recommended for connections of Best-Effort (BE) and Non-Real Time Variable Rate (NRT-VR) traffic. Under the sleep mode operation, sleep and listen windows are interleaved as long as
there is no downlink traffic destined to the node. During listen windows, the node checks with the base station whether there is any buffered downlink traffic destined to it in which case it leaves the sleep mode. Each sleep window is twice the size of the previous one but it is not greater than a specified final value. A node may awaken in a sleep window if it has uplink traffic to transmit.
\item
Type II classes are recommended for connections of Unsolicited Grant Service (UGS) and Real-Time Variable Rate (RT-VR) traffic. All sleep windows are of the same size as the initial window. Sleep and listen windows are interleaved as in type I classes. However, unlike type I classes, a node may send or receive traffic during listen windows if the requests handling time is short enough.
\item
Type III classes are recommended for multicast connections and management operations. There is only one sleep window whose size is the specified final value. At the expiration of this window, the node awakens automatically. 
\end{itemize}
The related operational parameters including the initial and maximum sleep window sizes can be negotiated between the mobile node and the base station. 

The sleep mode operation of IEEE 802.16e, more specifically the type I power saving class, has received an increased attention recently. In~\cite{vtc04}, the base station queue is seen as an $M/GI/1/N$ queueing system with multiple vacations; an embedded Markov chain models the successive (increasing in size) sleep windows. Solving for the stationary distribution, the dropping probability and the mean waiting time of downlink packets are computed. Analytical models for evaluating the performance in terms of energy consumption and frame response time are proposed in~\cite{Xiao05,Xiao06} and supported by simulation results. While~\cite{Xiao05} considers incoming traffic solely, both incoming and outgoing traffic are considered in~\cite{Xiao06}. In~\cite{vtc06}, the authors evaluate the performance of the type I power saving class of IEEE 802.16e in terms of packet delay and power consumption through the analysis of a semi-Markov chain. 

Power save mode in systems other than the IEEE 802.16e have also been studied; hereafter we cite some of these studies. In~\cite{chen99}, the authors evaluate the energy consumption of various access protocols for wireless infrastructure networks. The sleep mode operation of Cellular Digital Packet Data (CDPD) has been investigated through simulations in~\cite{cdpd99} and analytically in~\cite{cdpd03}. To efficiently support short-lived sessions such as web traffic, a bounded slowdown method -- that is similar to type I
power saving classes in the IEEE 802.16e -- is proposed for the IEEE 802.11 protocol in~\cite{hari02}. Last, the power saving mechanism for the 3G Universal Mobile Telecommunications System (UMTS) system is evaluated in~\cite{yang05}. 

In this paper, we propose a queueing-based modeling framework that is general enough to study many of the power save operations described in standards and in the literature. In particular, our model enables the characterization of the performance of type I and type II power saving classes as defined in the IEEE 802.16e standard~\cite{man}. The system composed of the base station, the wireless channel and the
mobile node is modeled as an $M/G/1$ queue with repeated inhomogeneous vacations. Traffic destined to the mobile node awaits in the base station as long as the node is in power save mode. When the node
awakens, the awaiting requests start being served on a first-come-first-served basis. The service consists of the handling of a frame at the base station, its successful transmission over the wireless channel and its handling at the node. Analytical expressions for the distribution and/or the expectation of many performance metrics are derived yielding the expected frame transfer time and the expected gain in energy. We formulate an optimization problem so as to maximize the energy efficiency gain, constrained to meeting some QoS requirements. We illustrate the proposed optimization scheme through four application scenarios.

Although we have motivated our modeling framework using power saving operation in wireless technologies, it is useful whenever the system can be modeled by a server with repeated vacations. The structure of the idle period is general enough to accommodate a large variety of scenarios. 

There has been a rich literature on queues with vacations, see e.g. the survey by Doshi(~\cite{Doshi}). Our model resembles the one of server with repeated vacations: a server goes on vacation again and again until it finds the queue non-empty. To the best of our knowledge, however, all existing models assume that the vacations are identically distributed whereas our setting applies to inhomogeneous vacations and can accommodate the case when the duration of a vacation increases in the average if the queue is found empty.

We exploit the well known tool "Stochastic decomposition" (\cite{Fuhrmann_Copper_1985}) to derive the main results.
Stochastic decomposition property of $M/G/1$ type queueing system with server vacation is one of the most remarkable results shown by \cite{Fuhrmann_Copper_1985}. The stationary queue length distribution at a random point in time can be decomposed in two or more parts where one part corresponds to stationary queue length distribution of M/G/1 system without vacation. This type of decomposition was first observed by \cite{Gaver_1962}, and subsequently by \cite{Miller_1964, Cooper_1970}, \cite{Levy_1975}, \cite{Shanthikumar_1980}, \cite{Scholl_Kleinrock_1983}, \cite{Ali_Neuts_1984}.
Considerable attention had been paid to the steady state analysis
under appropriate conditions on vacation sequences. Most of the
references can be found in two excellent review articles by
\cite{Doshi} and \cite{Teghem} in (1986).

Our model differs from the vacation model of \cite{Fuhrmann_Copper_1985}
due to the presence of inhomogeneous repeated vacation and warm-up time and vacation trigger time.
However, the decomposition property is still applicable to our
model since it holds the required assumptions stated
as in \cite{Shanthikumar}:
\begin{itemize}
\item The sequence of service times being independent of
arrival process and independent of sequence of vacation
periods that precede that service time. The service times
forms i.i.d. sequence.
\item Exhaustive service and the system is in queue is
stable, i.e. the server utilization is less than one.
More over packets do not balk, defect or renege from
the system.
\item Packets are served in First Come First Serve (FCFS) (or any order) which is
not dependent on their service times.
\item Service times is non preemptive.
\item Lack of Anticipation Assumption (LAA) for vacation termination.
\end{itemize}
Stochastic decomposition property allows us to obtain various
distribution using Probability generating function (pgf) and Laplace in section ~\ref{s:ana}) through highly
simplified approach which in turns yields more insight of the system.
The rest of the paper is organized as follows. Section~\ref{s:model}
describes our system model whose analysis is presented in
Sect.~\ref{s:ana}. Our modeling framework is applied to the power
saving mechanism in a WiMAX standard through four scenarios in
Sect.~\ref{s:illust}. Section~\ref{s:exploit} formulates several
performance and optimization problems whose results are shown and
discussed in Sect.~\ref{s:res}. Section~\ref{s:conc} concludes the
paper and outlines some perspectives.
\section{System Model and Notation}
\label{s:model}
\begin{table}

\centering
\begin{tabular}{p{0.10\columnwidth}p{0.8\columnwidth}}
\hline
{\it Symbol} & {\it Meaning}\\
\hline
\hline %
$N$ & Initial Queue size at the beginning of busy period ,\\
$X(t)$ & Queue size at time $t$ ,\\
$I$ & Idle duration in one cycle ,\\
$B$ & Busy duration in one cycle ,\\
$V_i$ & Size of $i$th Vacation ,\\
$\zeta$ & Expected number of vacations ,\\
$Q_N$ & Amount of work in queue with $N$ packets ,\\
${\cal L}_k(s)$ & Laplace of $k$th vacation with parameter $s$ ,\\
$T_w$ & Warm up duration (before busy period),\\
$T_{\tilde{w}}$ & Conditional warm up duration $T_{\tilde{w}}=T_w\bone\{t_f>T_t\}$,\\
$N(t)$ & Request arrival during time $t$ ,\\
$T_t$ & Vacation trigger time ,\\
$t_f$ & Time of first arrival if there is an arrival during $T_t$ ,\\
\hline
\end{tabular}\caption{Glossary : Main notation used throughout the paper}\label{tab:notation}
\end{table}

Consider an $M/G/1$ queue in which the server goes on vacation for a predefined period once the queue is observed empty for a vacation trigger duration. At the end of a vacation period, a new vacation initiates as long as no request awaits in the queue. We consider the exhaustive service regime, i.e., once the server has started serving customers, it continues to serve the queue until the queue empties. Request arrivals are assumed to form a Poisson process, denoted $N(t),t\geq0$, with rate $\lambda$. Let $\sigma$ denote a generic random variable having the same (general) distribution as the queue service times.

Note that the queue size at the beginning of a busy period impacts the duration of this busy period and is itself impacted by the duration of the last vacation period. Because arrivals are Poisson (a non-negative
L\'{e}vy input process would have been enough), the queue regenerates each time it empties and the cycles are i.i.d. Each regeneration cycle consists of:
\begin{enumerate}
\item {\em Vacation Trigger } time; Failing any arrival during trigger time, denoted by $T_t$, activate the vacation mode. However vacation is deferred if there is an arrival during $T_t$ which mimics the standard ${M/G/1}$ queue. Time of first arrival, denoted by $t_f$, is the idle duration , if the vacation is not triggered;
\item an {\em idle} period; let $I$ denote a generic random variable having
the same distribution as the queue idle periods, a generic idle
period $I$ consists of $\zeta$ vacation periods denoted $V_1,\ldots,V_\zeta$ and vacation trigger time $T_t$;
\item a {\em warm-up} period; it is a fixed duration denoted $T_{w}$ during which the server is warming up to start serving requests. $T_{\tilde{w}}$ denotes the conditional warm up duration when the vacation mode is triggered, i.e., $T_{\tilde{w}}=T_w\bone\{t_f>T_t\}$;
\item a {\em busy} period; let $B$ denote a generic random variable having
the same distribution as the queue busy periods. 
\end{enumerate}
The distribution of $V_i$ may depend on $i$, so the repeated vacations are {\em not} identically distributed. They are however assumed to be independent.

\begin{figure}[tb]
\centering
\scalebox{0.8}{ \input{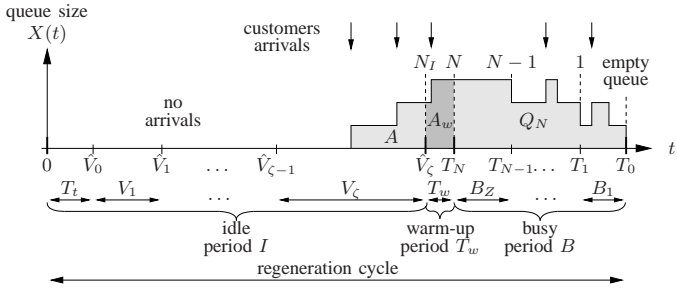} }
\caption{Sample trajectory of the queue size during a regeneration cycle.}
\label{fig:renewal} 
\end{figure}

Let $X(t)$ denote the queue size at time $t$. It will be useful to define the following instants relatively to the beginning of a generic cycle (in other words, $t=0$ at the beginning of the generic cycle):
\begin{itemize} 
\item $\hat V_i$ refers to the end of the $i$th vacation period, for $i=1,\ldots,\zeta$; observe that the idle period ends at $\hat V_\zeta$; we have $\hat V_i = \sum_{j=1}^i V_j$ and $I =\hat V_\zeta = \sum_{i=1}^{\zeta} V_i$; \item $T_N$ refers to the beginning of the busy period $B$; we define
$N:=X(T_N)$ as the queue size at the beginning of a busy period; 
\item $T_i$ refers to the first time the queue size {\em decreases} to the value $i$ (i.e. $X(T_i)=i$) for $i=N-1,\ldots,0$; observe that the cycle ends at $T_0$.
\end{itemize}
The times $\{T_i\}_{i=N,N-1,\ldots,0}$ delimit $N$ sub periods in $B$, as can be seen in Fig.~\ref{fig:renewal}. We can write $B=\sum_{i=1}^NB_i$ where $B_i=T_{i-1}-T_i$.
The random variable $N$ is in fact the number of arrivals from $t=0$ until time $T_N$, even though all of the arrivals occur between $\hat V_{\zeta -1}$ and $T_N$. Introduce $N_I$ as the number of requests that have arrived up to time $\hat V_\zeta$ (i.e. during period $I$) and $N_{T_w}$ as the number of arrivals during the warm-up period $T_w$. Hence $N=N_I+N_{T_w}$. Note that the queue size at the end of idle duration is $X(I)=N_I$. 

A possible trajectory of $X(t)$ during a regeneration cycle is depicted in Fig.~\ref{fig:renewal} where we have shown the notation introduced so far. 

\section{Analysis}
\label{s:ana}
This section is devoted to the analysis of the queueing system presented in Sect.~\ref{s:model}. We will characterize the distributions of $\zeta$ and $N$, derive the expectations of $\zeta$, $I$, $N$, $B$ and $X(t)$ and the second moments of $I$ and $N$, and lastly compute the system response time. The gain from idling the server is introduced in the special case when the model is applied to study the power save operation in wireless technologies; see Sect.~\ref{s:illust}.

\subsection{The Number of Vacations}
To compute the distribution of $\zeta$, the number of vacation periods during an idle period, we first observe that the event $\zeta\geq i$ is equivalent to the event of no arrivals during $\hat V_{i-1}=\sum_{k=1}^{i-1}V_k$. Note that $\zeta \geq 1$ reflects that the first arrival occurred after the trigger wait time $T_t$. Equivalently, the event $\zeta=0$ reflects that the at least one arrival occurred during the vacation trigger time $T_t$. When $\zeta=0$, let the arrival time of the first customer is denoted by $t_f$
conditioned such that $t_f<T_t$. 

Let $A_{T_t}$ denote the event of no arrival during $[0,T_t]$ and ${\cal L}_{T_t}(\lambda)$ denotes  $e^{-\lambda T_t}$. Let $A_k$ denote the event of no arrivals during the period of time $V_k$, and $A_k^c$ denote the complementary event. Denoting by ${\cal L}_k(s):=\E[\exp(-s V_k)]$ and ${\cal L}_{\widehat{i-1}}(s):=\E[\exp(-s \hat{V}_{i-1})]$ the Laplace Stieltjes Transform (LST) of $V_k$ and $\hat V_{i-1}$ respectively, we can readily write \vspace{-1mm}
\bear
P(\zeta=0)&=&P(A_{T_t}^c)=\E[\bone\{A_{T_t}^c\}]=\E[\E[\bone\{A_{T_t}^c\}|T_t]]
\nonumber\\ 
\label{e:pzeta-0}&=& \E\left[1-\exp(-\lambda T_t)\right]=1-{\cal L}_{T_t}(\lambda),\\
P(\zeta=1)&=&P(A_1^c)=\E[\bone\{A_1^c\}]=\E[\E[\bone\{A_1^c\}|V_1]]
\nonumber\\
&=&\nonumber\E\left[\exp(-\lambda T_t)\right]\E\left[1-\exp(-\lambda V_1)\right]\\
&&={\cal L}_{T_t}(\lambda)(1-{\cal L}_1(\lambda)),
\label{e:pzeta-1}
\eear
and for $i>1$, we have\vspace{-1mm}
\bear
P(\zeta=i)|_{(i>1)}&=&P(A_{T_t})\prod_{k=1}^{i-1}P(A_k)\,P(A_i^c)\nonumber\\
&&\hspace{-20mm}={\cal L}_{T_t}(\lambda)\left(\prod_{k=1}^{i-1}L_k(\lambda)\right)(1-{\cal L}_i(\lambda)),
\label{e:pzeta-i}\\
\nonumber P(\zeta\geq i)|_{(i>1)}&=&P(A_{T_t})\prod_{k=1}^{i-1}P(A_k)\\
&&\hspace{-20mm}={\cal L}_{T_t}(\lambda)\prod_{k=1}^{i-1}L_k(\lambda)
= {\cal L}_{T_t}(\lambda){\cal L}_{\widehat{i-1}}(\lambda), \label{e:pzeta>i} \eear
where we have used the fact that arrivals are Poisson with rate $\lambda$. The product $\prod_{k=a}^b{\cal L}_k(\lambda)$ is defined as equal to 1 for any $b<a$. Let ${\cal L}_{T_t}(s):=\exp(-s
T_t)$. Using~\eqref{e:pzeta>i}, the expected number of vacations in an idle period is given by
\beq 
\E[\zeta]=\sum_{i=0}^{\infty}iP(\zeta=i)=\sum_{i=1}^{\infty}P(\zeta\geq i)
={\cal L}_{T_t}(\lambda)\sum_{i=1}^{\infty}{\cal L}_{\widehat{i-1}}(\lambda).
\label{e:Ezeta}
\eeq

\subsection{The Idle Period}
The system goes on vacation only if the inactivity duration is more than the trigger time $T_t$. Therefore, if the vacation is triggered the idle period is the sum of all the vacation durations including vacation trigger time. Otherwise, Idle period is only the duration of first arrival ( idle period of standard $M/G/1$ queue). 
The idle period is thus given by 
\bear
I&=&\Big[ T_t+\sum_{i=1}^\zeta V_i\Big]{\bone^c{\{\zeta=0\}}} +{T_f}{\bone{\{\zeta=0\}}}
\nonumber \\&&= \min[T_t, t_f]+\sum_{i=1}^\zeta V_i{\bone{\{\zeta\neq 0\}}}.
\eear

Using the equality $\sum_{i=1}^{\zeta} V_i=\sum_{i=1}^{\infty}
V_i{\bone {\{\zeta \geq i\}}}$, expected idle period can be obtained
as follows
\bear
\nonumber \E[I]&=&\E\left[\min(T_t,t_f)\right]+\E\left[\left(\sum_{i=1}^{\infty} V_i{\bone {\{\zeta \geq i\}}}\right){\bone {\{\zeta\neq 0\}}}\right]\\
\nonumber &=&\E[(T_f)]\P(\zeta=0)+\E[T_t]\P(\zeta \neq 0)\\
&&\nonumber + \left[\sum_{i=1}^{\infty} \E [V_i e^{-\lambda(\hat{V}_{i-1})} ]{\cal L}_{T_t}(\lambda)\right]\\
\label{e:EI} &&\hspace{-10mm}=\frac{1}{\lambda}{\cal L}^c_{T_t}(\lambda)+T_t {\cal L}_{T_t}(\lambda)+ {\cal L}_{T_t}(\lambda)\sum_{i=1}^{\infty}\E[V_i] {\cal L}_{\widehat{i-1}}(\lambda)\hspace{4mm}
\eear
Note that when $T_t=0$, the idle period reduces to
$\E[I]=\sum_{i=1}^{\infty} \E [V_i]\prod_{k=1}^{i-1} {\cal L}_k(\lambda)$, which
is in congruence with \cite{Qest08}(eq.4).
\subsection{The Initial Queue Size Distribution in Busy Period}
The number of requests/packets waiting in the queue at the beginning of busy period is $N=N_I+N_{T_{\tilde{w}}}$, where $T_{\tilde{w}}=T_{w}(\bone\{(t_f>T_t)\})$. The indicator function simply indicates that the warm up period is accounted only when sleep mode is triggered. Since the number of arrival during the idle period is independent of the arrival during the warm up period $T_w$,
the pgf of $N$ is the product of pgf of $N_I$ and $N_{T_w}$. Therefore, we have the pgf of $N$  given as 
\bear N(z)=N_I(z)~N_{T_{\tilde{w}}}(z). \eear
The queue size pgf during the idle period $N_I(z)$ is given by
\bear \label{e:ZI_z} N_I(z)=\sum_{i=0}^\infty
z^i~\P(N_I=i)=\sum_{i=1}^\infty z^i~\P(N_I=i).
\eear
The last equality above is due to the fact that at least one arrival is sure during idle period, i.e., $\P(N_I=0)=0$. The number of arrivals during the entire idle period is the sum of arrivals during each vacation periods $V_i$'s if there are any. To derive the distribution of $N_I$, we first compute the joint distribution of $N_I$ and $\zeta$, the number of vacations in an idle period. Observe that $N_I$ takes value in $\N^*$. Noting that the vacation trigger event $\zeta=0$ is also possible, we can write 
\bears
&&P(N_I=j,\zeta=i,\zeta \geq 1)
\\&&=P\Big(j \textrm{ arrivals in }V_i,\bone\{A_1,\ldots,A_{i-1}\}, \bone^c\{A_{T_t} \} \Big)\\
&=&\E\left[\exp(-\lambda V_i)\frac{(\lambda V_i)^j}{j!}\right]\prod_{k=1}^{i-1} {\cal L}_k(\lambda){\cal L}_{T_t}(\lambda).
\eears
and, 
\bears
{P(N_I=j,\zeta = 0)}
= \left\{
\begin{array}{ll}
1-{\cal L}_{T_t}(\lambda), & \hbox{if } j = 1 \\
0, & \mbox{otherwise.}
\end{array}
\right.
\eears
The first term refers to the scenario when vacation is triggered, while the second term depicts when vacation is not triggered. In that case there is an arrival before the vacation trigger time $T_t$.
Therefore, in that case it is nothing but a standard $M/G/1$ queue (without vacation).

Denoting ${\cal L}^c_{T_t}(\lambda))=(1-{\cal L}_{T_t}(\lambda))$, the z transform of initial queue size $Z(.)$ is given using eq. \eqref{e:ZI_z} by
\bear
\nonumber N_I(z)&=&\sum_{m=0}^{\infty}z^m \P(N_I=m)=z\P(N_I=1)\\
&&\nonumber +\sum_{m=2}^\infty z^m \P(N_I=m) \\
&=&{\cal L}^c_{T_t}(\lambda) z+\sum_{i=1}^{\infty} {\cal
L}_{i}(\lambda(1-z)) {\cal
L}_{\widehat{i-1}}(\lambda){\cal L}_{T_t}(\lambda). \hspace{6mm}\eear
Since the arrival is a Poisson process, the pgf of arrival during the fixed warm up period $T_w$ is given as
\bear N_{T_{\tilde{w}}}(z)\nonumber&=&\sum_{i=0}^\infty[\P(t_f>T_t)z^i\P(N_{T_{\tilde{w}}}=i)\\
&&\nonumber+\P(t_f\leq T_t)z^i\P(N_{T_{\tilde{w}}}=i)]\\
&=& {\cal L}_{T_t}(\lambda){\cal
L}_{T_w}(\lambda(1-z))+ {\cal L}^c_{T_t}(\lambda).\eear
Where the Laplace transform of the arrivals during the warm up period $T_w$ is
$
N_{T_{{w}}}(z)=
e^{-\lambda T_w(1-z)}:={\cal L}_{T_w}(\lambda(1-z)). $

The above equations combines to yield $N(z)=N_I(z)N_{T_{\tilde{w}}}(z)$, given by
\bear \label{e:Zz}
\nonumber N(z)&=&\left(z {\cal L}^c_{T_t}(\lambda)+\sum_{i=1}^{\infty}
{\cal L}_{i}(\lambda(1-z)) {\cal
L}_{\widehat{i-1}}(\lambda){\cal L}_{T_t}(\lambda)\right)\\&&~\Big({\cal L}_{T_t}(\lambda){\cal
L}_{T_w}(\lambda(1-z))+ {\cal L}^c_{T_t}(\lambda)\Big).\eear
%
Setting $T_t=0$ corresponds the model presented in \cite{Qest08} which is forced vacation scenario, while $T_t=\infty$ corresponds to simple the M/G/1 queue. 

Noting that, $z$ transform is one of well known tool to obtain moments by using the relation $N^{(n)}(1)=\E[N(N-1)\ldots (N-i+1)]$, which simply means the evaluation of $n$th derivative of $N$, denoted as $N^{(n)}(,)$, at $z=1$, we derive the first, second and third derivatives of $N(z)$ (which
will be required in latter sections). The first derivative of $N(z)$ is given by
\bear \label{e:Nzd1}
N^{(1)}(z)&=&N_I(z)N^{(1)}_{T_{\tilde{w}}}(z)+N^{(1)}_I(z)N_{T_{\tilde{w}}}(z).
\eear %
where 
\bear
\label{e:N1Tw}
N^{(1)}_{T_{\tilde{w}}}(z)&=& {\cal L}_{T_t}(\lambda)\lambda T_w {\cal L}_{T_w}(\lambda(1-z)),\\
\label{e:ZI'z}
N^{(1)}_I(z)&=&{\cal L}^c_{T_t}(\lambda)+ \sum_{i=1}^{\infty} \frac{d{\cal L}_{i}(\lambda(1-z))}{dz} {\cal L}_{\widehat{i-1}}(\lambda){\cal L}_{T_t}(\lambda).\hspace{6mm}
\eear
From the definition of $z$ transform we know that $N_{T_{\tilde{w}}}(1)=N_I(1)=1$ and thus $N^{(1)}_{T_{\tilde{w}}}(1)=\lambda T_w {\cal L}_{T_t}(\lambda)$. Further substituting eq. \eqref{e:EI} in $N^{(1)}_I(.)$, we have
\bear
\label{e:EZ}
\nonumber \E[N_I]&=& 
\lambda\left[ \frac{{\cal L}^c_{T_t}(\lambda)}{\lambda}+{\cal L}_{T_t}(\lambda)\sum_{i=1}^{\infty}
\E[V_i]{\cal L}_{\widehat{i-1}}(\lambda) \right]\\
&&=\lambda\left[\E[I]-T_t {\cal L}_{T_t}(\lambda)\right]=\lambda\E[\widetilde{I}]
\eear
where $\E[\widetilde{I}]:=\E[I]-T_t {\cal L}_{T_t}(\lambda)$.
Combining all together we obtain the first moment,
\bear
\nonumber
\E[N]&=&N^{(1)}(1)=N_I(1)N^{(1)}_{T_{\tilde{w}}}(1)+N^{(1)}_I(1)N_{T_{\tilde{w}}}(1)\\
\label{eq:EN}&=&\lambda T_w {\cal L}_{T_t}(\lambda)+\lambda \E[\widetilde{I}].
\eear
\noindent Proceeding further for the second moment, the derivative of eq. \eqref{e:Nzd1} is 
\bear
\nonumber N^{(2)}(z)&=& N_I(z)N^{(2)}_{T_{\tilde{w}}}(z) +N_{T_{\tilde{w}}}(z) N^{(2)}_I(z)\\
&&\label{e:Z''z}+ 2 N^{(1)}_{T_{\tilde{w}}}(z) N^{(1)}_I(z).\eear
where, using eq. \eqref{e:ZI'z} and eq. \eqref{e:N1Tw}, we have
\bear
\nonumber N^{(2)}_I(z)&=&\sum_{i=1}^{\infty} \frac{d^2{\cal L}_{i}(\lambda(1-z))}{dz^2} {\cal L}_{\widehat{i-1}}(\lambda){\cal L}_{T_t}(\lambda),\\
\label{e:ZI''z} N^{(2)}_{T_{\tilde{w}}}(z)&=& (\lambda T_w)^2 {\cal L}_{T_t}(\lambda) {\cal L}_{T_w}(\lambda(1-z)).
\eear
Evaluating at $z=1$, we have
\bear
\label{e:EIa1}
&&\hspace{-9mm}N^{(2)}_I(1)=\sum_{i=1}^{\infty} \E[(\lambda V_i)^2]{\cal L}_{\widehat{i-1}}(\lambda){\cal L}_{T_t}(\lambda) =\lambda^2 \E[I_a], \hspace{2mm}\\
&&\hspace{-9mm}N^{(2)}_{T_{\tilde{w}}}(1)=(\lambda T_w)^2 {\cal L}_{T_t}(\lambda).
\eear
where $E[I_a]:=\sum_{i=1}^{\infty} \E[ V_i^2]{\cal L}_{\widehat{i-1}}(\lambda){\cal L}_{T_t}(\lambda)$.
Thus, from eq. \eqref{e:Z''z} we can write
\bear
\label{e:Z''1}
\hspace{-2mm}N^{(2)}(1) = \lambda^2\left(T_w^2 {\cal L}_{T_t}(\lambda)+2 T_w {\cal L}_{T_t}(\lambda)\E[\widetilde{I}]+ \E[I_a]\right).
\eear
Therefore the second moment is
\bear
\nonumber &&\hspace{-10mm}\E[N^2]=\E[N]+N^{(2)}(1)=\lambda({\cal L}_{T_t}(\lambda)T_w+\E[\widetilde{I}])\\
\label{eq:EN2}&&+\lambda^2({\cal L}_{T_t}(\lambda)T_w^2+2 T_w{\cal L}_{T_t}(\lambda)\E[\widetilde{I}]+\E[I_a]).
\eear
Note that expected initial queue size $\E[N]$ and its second moment $\E[N^2]$ obtained in \cite{Qest08}[eq. 11,12] is a special case of eq. \eqref{eq:EN}, \eqref{eq:EN2} when $T_t=0$ (forced vacation scenario). Note that, setting the parameter $T_t=0$ essentially means that the vacation must be triggered whenever the system is idle, which is a particular case.

\subsection{Queue Size Distribution}
Queue size is given by the number of requests/packets in the queue seen by any random arriving packet. From "Poisson Arrival See Time Average" (PASTA) (\cite{Pasta_1982}) the stationary queue size is equivalent to the number of requests waiting in the queue to be served left behind by a random departure. Further more, from \cite{Kleinrock}, it is equivalent to the queue size using the workload decomposition or stochastic decomposition approach (refer to \cite{Fuhrmann_Copper_1985}). The pgf of the stationary distribution of the number of request/packets left behind that a random departing customer is given by
\bear
\label{eq:decom1}
X(z)=\frac{1- N(z)}{\E[N](1-z)} X_{M/G/1}(z).
\eear
\noindent
where $Z(.)$ denotes the pgf of queue length at
the beginning of busy period, $X_{M/G/1}(.)$ denotes the pgf of the number of customer left behind in a standard $M/G/1$ queue. In stationary regime, the distribution of queue length can also be
given by $X(z)$. From \cite{Cooper_book}(pp. 210), the pgf of a standard $M/G/1$ queue is given by
\bear
\label{e:Xz}
X_{M/G/1}(z)=\frac{(1-\rho)(1-z)\sigma(\lambda-\lambda z)}{\sigma(\lambda-\lambda z)-z}~.
\eear
where $\sigma(.)$ is the service time distribution.%
\subsubsection{Expected Queue length}
The moments of queue length can be directly obtained from $X(z)$ by calculating its derivatives at $z=1$.
We double derivate eq. \eqref{eq:decom1} with respect to $z$ and do some simple calculus to obtain
the expectation of $X(.)$ which is given by 
\bear
&&\nonumber\E[N]\left((1-z)X^{(2)}(z)-2X^{(1)}(z)\right)\\
&&\nonumber =(1-N(z))X^{(2)}_{M/G/1}(z)-2N^{(1)}(z)X^{(1)}_{M/G/1}(z)\\
\label{e:Z''}&&\hspace{10mm}-N^{(2)}(z)X_{M/G/1}(z).
\eear
Using the relations
$[1-Z(1)]=0$ (which is direct from eq. \eqref{eq:decom1}) at $z=1$,
we have
\bear
\E[X]=X^{(1)}(1)=\frac{N^{(2)}(1)}{2\E[N]}+\E[X_{M/G/1}]~.
\eear
Substituting
$N^{(2)}(1)$ from \eqref{e:Z''1} and
$X^{(1)}_{M/G/1}(1)=\E[X_{M/G/1}]= \rho+\frac{\lambda^2\E[\sigma^2]}{2(1-\rho)}$
from \cite{Kleinrock} (Pollaczek-Khinchin mean value formula(sec.
5.6)) in eq. \eqref{e:Z''},
we finally obtain the expected queue length

\bear
\label{e:EX}
\E[X]&=&\frac{\lambda(T_w^2{\cal L}_{T_t}(\lambda)+2T_w{\cal L}_{T_t}(\lambda)\E[\tilde{I}]+\E[I_a])}{2(T_w{\cal L}_{T_t}(\lambda)+\E[\tilde{I}])}
\nonumber\\
&&+\left(\rho+\frac{\lambda^2\E[\sigma^2]}{2(1-\rho)}\right) .
\eear

\subsection{Busy period}
The length of the busy period, denoted by $B$, depends on the number of customers/packets are waiting at the end of the vacation interval. If there are $N$ packet requests are waiting, the subsequent busy period will consists of $N$ independent busy periods, each of which is denoted by $B_1$ which mimics the single request service time as in $M/G/1$ queue. Therefore, we have 
\bears
{\cal B}^*(s)= \sum_{i=1}^\infty N_I[{\cal B}_1^*(s)]^i=N[{\cal B}_1^*(s)]
\eears
Thus the expected busy period can be given by
\bear
\E[{\cal B}]=\E[N]\E[{\cal B}_1]=\frac{\E[N]\E[\sigma]}{1-\rho}
\eear
\subsection{Sojourn time}
Assume the waiting time of a customer is independent of the part of the arrival process that occurs after the customer's arrival epoch, which is easy to show. Our policy, which is FCFS discipline, falls in this category. The waiting time of an arbitrary customer in a queue is exactly the number of customer ahead of the tagged customer in the queue under FCFS scheme. The number of customers left behind by the tagged customers is precisely the number of arrivals during the sojourn time (waiting + service) of the tagged customers, denoted its pgf by $N(z)$. Since Poisson arrival see time average (PASTA, see Wolf), the pgf of number customer ahead of a random customer has the same pgf as $N(z)$. Therefore we can express the LST of the waiting time $W^*(s)$ of a random customer in the queue as (from \cite{Fuhrmann_Copper_1985}) 
\bear
W^*(s)&=& \frac{\lambda[1-N(1-s/\lambda)]}{s \E[N]} W^*_{M/G/1}(s).
\eear
Where, $W^*_{M/G/1}(s)$ is the LST of waiting time of an arbitrary request in the queue (excluding its service time) of a standard $M/G/1$ queue. From \cite{Takagi}(1.45), we have
$W^*_{M/G/1}(s)=\frac{s(1-\rho)}{s-\lambda+\lambda \sigma^*(s)}$, where
LST of service time is given by $\sigma^*(s)=\E[e^{-s\sigma}]$.
Thus, we have
\bear
\nonumber
W^*(s)&=& \frac{\lambda[1-N(1-s/\lambda)]}{s \E[N]} \frac{s(1-\rho)}{s-\lambda+\lambda \sigma^*(s)}\\
\label{e:Ws}&=& K \frac{[1-N(1-s/\lambda)]}{s-\lambda+\lambda \sigma^*(s)}.
\eear
where $K=\frac{(1-\rho)\lambda}{\E[N]}$.
The moments of the $W(.)$ can be obtained from its LST by simply evaluating its derivatives at $s=0$, i.e., $E[W^{*n}]=(-1)^nW^{*{(n)}}(0)$ as follows, (Refer appendix \ref{sec:a_sojourn_time} for detailed computation),
\bear
\E[W]=\frac{{N^{(2)}}(1)}{2\lambda\E[N]}+ \frac{\lambda\E[\sigma^2]}{2(1-\rho)}~.
\eear

\subsection{Message Response Time}
The {\em message response time} $T$ is defined as the time interval from the arrival time of an arbitrary message to the time when it leaves the system after the service completion. The mean message response time is said to be the \emph{single most important performance measure} \cite{Kleinrock}(page 162) for the system without blocking.

The response time of a message consists of the { \em waiting time} $W$ and the service time $\sigma$. Since the waiting time and service times are independent, we can express the LST of response time $T$ and mean waiting time directly as follows
\bear
T^*(s)=W^*(s)\sigma^*(s) \Rightarrow \E[T]=\E[W]+\E[\sigma].
\eear

Thus, we can express
\bear
\nonumber
\E[T]&=& \frac{{N^{(2)}}(1)}{2\lambda \E[N]}+\frac{\lambda \E{\sigma^2}}{2(1-\rho)}+\E[\sigma]\\
&=&\label{e:T-alt}\frac{T_w^2{\cal L}_{T_t}(\lambda)+2T_w{\cal L}_{T_t}(\lambda)\E[\tilde{I}]+\E[I_a])}{2(T_w{\cal L}_{T_t}(\lambda)+\E[\tilde{I}]}\nonumber \\
&&+\left(\frac{\lambda \E[\sigma^2]}{2(1-\rho)}+\E[\sigma]\right).
\eear
The last term in bracket above is the mean response time of standard $M/G/1$ queue, denoted by $\E[T_{M/G/1}]$, (refer \cite{Takagi}), while the first part is the additional contribution due to vacation(which includes warm up and trigger time). Therefore, the expected sojourn time can be rewritten as
\bear
\label{e:ET}
\E[T]=\frac{T_w^2{\cal L}_{T_t}(\lambda)+2T_w{\cal L}_{T_t}(\lambda)\E[\tilde{I}]+\E[I_a]}{2(T_w{\cal L}_{T_t}(\lambda)+\E[\tilde{I}])}+\E[T_{M/G/1}]
\eear
\begin{remark}
One can also obtain the expected time a customer spends in the queue using Little's formula as follows,
\beq
\label{e:T}
\E[T]=\frac{\E[X]}{\lambda},
\eeq
On substitution of $\E[X]$ from \eqref{e:EX}, we obtain the same expression of \eqref{e:ET}.
\end{remark}

As the rate $\lambda\to1/\E[\sigma]$ (recall that the stability condition enforces that $\lambda\E[\sigma]<1$), we must have $P(\zeta=1)\to1$ (thus ${\cal L}_1(\lambda)\to0$) whatever the distribution of the vacations. There will then be only one vacation period in most idle periods. Therefore, at large input rates, the largest contribution to the sojourn time is expected to come from the waiting time when the server is active (queueing delays).

\subsection{ Excess Waiting time }
\label{sec:a_excess}
Excess waiting time could be a performance metric of interest in context of QoS of delay sensitive class of traffic. In several applications (e.g. VoIP), it is important to observe the excess waiting time. A packet arriving late then certain delay is not useful and may be required to discard. Motivated from such necessity, we briefly discuss the excess waiting time. The main idea is to state that we can estimate the bound on excess waiting time by exploiting Markov inequality (\cite{nelson}[sec. 5.5.1]). We obtain the bound on the probability of waiting time as follows using first and second moments as follows:
\bears
\P(W>w)\leq \frac{\E[W]}{w}, \quad \P(W>w)\leq \frac{\E[W^2]}{w^2}
\eears
Let us denote $M_1=\frac{\E[W]}{w}$, and $M_2=\frac{\E[W^2]}{w^2}$.
Thus we have,
\bear
\nonumber &&\hspace{-7mm}M_2=\frac{\E[W^2]}{w^2},\\
&&\hspace{-7mm}=\frac{ T_w^3 {\cal L}_{T_t}(\lambda)+\E[I_c]+3 (T_w^2{\cal L}_{T_t}(\lambda)\E[\widetilde{I}] +{\cal L}_{T_t}(\lambda) \E[I_a])}{3w^2(T_W{\cal L}_{T_t}(\lambda)+\E[\widetilde{I}])}\nonumber\\
&&\hspace{-7mm}+
\frac{\lambda \E[\sigma^{2}]}{w^2(1-\rho)}\E[W] +\frac{\lambda \E[\sigma^{3}]}{3w^2(1-\rho)},\eear
\bear
&&\hspace{-10mm}M_2=\frac{ T_w^3 {\cal L}_{T_t}(\lambda)+\E[I_c]+3 (T_w^2{\cal L}_{T_t}(\lambda)\E[\widetilde{I}] +{\cal L}_{T_t}(\lambda) \E[I_a])}{3w^2(T_W{\cal L}_{T_t}(\lambda)+ \E[\widetilde{I}])}\nonumber\\
&&\hspace{-7mm}+
\frac{\lambda \E[\sigma^{2}]}{w(1-\rho)}M_1 +\frac{\lambda \E[\sigma^{3}]}{3w^2(1-\rho)}.
\eear

Clearly, we can state that if $\frac{\lambda \E[\sigma^2]}{(1-\rho)w}>1$, then $M_1<M_2$( however vice versa may not be true). This suggest that $M_1$ is tighter than $M_2$ under condition $\frac{\lambda
\E[\sigma^2]}{(1-\rho)w}>1$. 
The bound can be utilized to approximate the distribution of excess waiting time. For example, we can easily estimate the maximum packet dropping probability for limited buffer capacity scenario. We do not aim to derive a tight bound for excess waiting time distribution, rather we just illustrate that 
many of the derived analytic results can be easily extended to approximate the distributions of various performance entities, e.g. service delay, dropping probabilities.

\section {Application to Power Saving}
\label{s:illust}
The model analyzed in Sect.~\ref{s:ana} can be used to study energy saving schemes used in wireless technologies. Consider the system composed of the base station, the wireless channel and the mobile node. When the energy saving mechanism is disabled, the system can be seen as an $M/G/1$ queue; and when it is enabled, the system can be modeled as an $M/G/1$ queue with vacations. The server goes on vacations repeatedly until the queue is found non-empty. This models the fact that the mobile node goes to sleep by turning off the radio as long as there are no packets destined to it.

In practice, the mobile needs to turn on the radio to check for packets. The amount of time needed is called the {\em listen} window and is denoted $T_l$. During a listen window, the mobile can be informed of any packet that has arrived {\em before} the listen window. Any arrival during a listen window can only be notified in the following listen window. To comply with this requirement, we will make all but the first vacation periods start with a listen window $T_l$. The last listen window is included in the warm-up period $T_w$ ( $T_w=T_l$ is considered for numerical evaluation).

\begin{figure}[h]
\begin{center}
\subfigure[ Normal mode of a mobile node.]{
\label{f:map}
\resizebox{.42\textwidth}{.15\textwidth}{\input{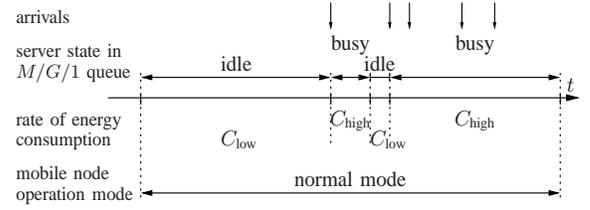}}}
\subfigure[Repeated vacations to the possible states of a mobile node.]{
\label{fig:map}
\resizebox{.42\textwidth}{.19\textwidth}{ \input{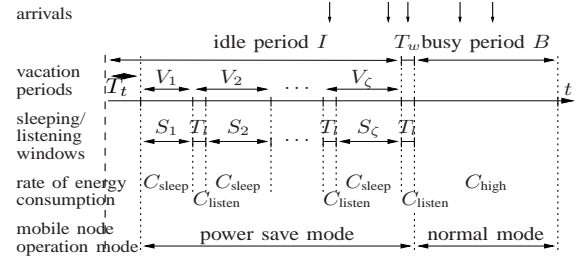}}}
\caption{Mapping the $M/G/1$ queue .}
\end{center}{}
\end{figure}

Let $S_i$ be a generic random variable representing the time for which a node is sleeping during the $i$th vacation period. We then have $V_1=S_1$ and $V_i=T_l+S_i$ for $i=2,\ldots,\zeta$. In this paper, we are assuming $T_l$ to be a constant. As for the $\{S_i\}_i$, four cases will be considered as detailed further on. Figure~\ref{f:map} (resp.~\ref{fig:map}) maps the state of an $M/G/1$ queue (resp. an $M/G/1$ queue with repeated vacations) to the possible states of a mobile node.

\subsection{The Energy Gain under Power Saving}
The performance metric defined in this section complements the ones derived in Sect.~\ref{s:ana}, but is specific to applications in wireless networks, and more precisely, to energy saving mechanisms. In
this section, we will derive the gain in energy at a node should the power save mechanism be activated.

Having noted the possible node states, we can distinguish between four possible levels of energy consumption, that are, from highest to lowest,
\begin{itemize}
\item $C_{\textrm{high}}$: experienced during exchanges of packets which includes the busy period($B$),
\item $C_{\textrm{listen}}$: experienced when checking for downlink packets which includes the listening periods $T_l$,
\item $C_{\textrm{low}}$: the lowest level observed when the mobile
node is inactive but not in sleep state which includes the duration $I-\sum_{i=1}^\zeta V_i$,
\item $C_{\textrm{sleep}}$: the lowest level observed when the mobile
node is in sleep state which includes the total vacation duration given by $\sum_{i=1}^\zeta V_i$.
\end{itemize}
When the power save mechanism is not activated, the energy consumption per unit of time is $C_{\textrm{low}}$ in idle periods (whose expectation is $1/\lambda$) and is equal to $C_{\textrm{high}}$ during the busy periods (whose expectation is $\E[B_1]$). The energy consumption rate can be written
\beq
\label{e:Enosleep}
E_{\textrm{no sleep}} := \rho\, C_{\textrm{high}}+(1-\rho) C_{\textrm{low}}
\eeq
where $\rho=\lambda\E[\sigma]=\E[B_1]/(1/\lambda+\E[B_1])$ (loss free system).

Consider now the case when the power save mechanism is activated. During busy periods, \bear
\nonumber
E_{\textrm{sleep}}&:=&\frac{1}{{\E[C]}} \left(\E[I\bone\{t_f>T_t\}]C_{\textrm{sleep}}-\E[I\bone\{t_f\leq T_t\}\right.\\
&&\left. C_{\textrm{low}}]+ T_t\E[\bone\{t_f>T_t\}](C_{\textrm
{low}}-C_{\textrm
{sleep}})\right.\nonumber\\
&&\left.+\E[T_l(\zeta-1)\bone\{t_f>T_t\}](C_{\textrm{listen}}-C_{\textrm{sleep}}) \nonumber\right.\\
&&\left.+\E[T_w\bone\{t_f>T_t\}]C_{\textrm{listen}}
+\E[B]C_{\textrm{high}} \right)
\label{e:Esleep}
\eear
where the average cycle duration can be computed as \small{ $\E[C]:=\E[I+T_w\bone\{t_f>T_t\}+B]=\E[I]+T_w{\cal L}_{T_t}(\lambda)+\E[B]$.}
Evaluating the above terms we obtain
\bear
\nonumber
&&\nonumber \hspace{-8mm}E_{\textrm{sleep}}=\frac{ (\E[I]-\E[\tilde{I}])C_{\textrm{sleep}}+\E[\tilde{I}]C_{\textrm{low}} +T_w {\cal L}_{T_t}C_{\textrm
{listen}} +\E[B]C_{\textrm{high}}}{\E[C]}\\&&+\frac{\E[\zeta-1]T_l{\cal L}_{T_t}(\lambda) (C_{\textrm{listen}}-C_{\textrm{sleep}})}{\E[C]}
\label{e:Esleep1}
\eear
where $\E[I]$ is defined in eq. \eqref{e:EI} and denote $\E[\tilde{I}]:=\frac{1}{\lambda}{\cal L}_{T_t}(\lambda)$.
Note the following,
\bear
\nonumber
&&\hspace{-8mm}\E[I\bone\{t_f\leq T_t\}]=\E[\min(T_t,t_f)\bone\{t_f\leq T_t\}]\\&&\hspace{-4mm}+\E\left[\left(\sum_{i=1}^{\infty} V_i{\bone {\{\zeta \geq i\}}}\right){\bone {\{\zeta\neq 0\}}}]\bone\{t_f\leq T_t\}\right]
\E[t_f] {\cal L}_{T_t}(\lambda)\nonumber\\
&&\hspace{15mm}:=\E[\tilde{I}].
\eear
and
\bear
\nonumber
\E[I\bone\{t_f>T_t\}]&=&\E[\min(T_t,t_f)\bone\{t_f>T_t\}]\\
&&\hspace{-25mm}+\E\left[\left(\sum_{i=1}^{\infty} V_i{\bone {\{\zeta \geq i\}}}\right){\bone {\{\zeta\neq 0\}}}]\bone\{t_f>T_t\}\right]
T_t {\cal L}_{T_t}(\lambda)\nonumber \\&&\hspace{-25mm}+E\left[\sum_{i=1}^{\infty} V_i\right] {\cal L}_{T_t}(\lambda)
:=\E[I]-\E[\tilde{I}].
\eear

Observe that $\E[B]/\E[C]=\rho=\lambda\E[\sigma]$ because  we have assumed an unlimited queue (no overflow losses). The economy in energy per unit of time should a node enable its power saving mechanism is  $E_{\textrm{nosleep}}-E_{\textrm{sleep}}$. The relative economy, or the {\em
energy gain} is defined as 
\bear
\label{e:G-def}
G&:=&\frac{E_{\textrm{no sleep}}-E_{\textrm{sleep}}}{E_{\textrm{no sleep}}} \\
\nonumber&=&
\frac{(1-\rho)\frac{C_{\textrm{low}}}{C_{\textrm{high}}}}{\rho+(1-\rho)\frac{C_{\textrm{low}}}{C_{\textrm{high}}}}
-\frac{\rho/\E[B]}{({\rho+(1-\rho)\frac{C_{\textrm{low}}}{C_{\textrm{high}}}})}\\
&&
\left((\E[I]-\E[\tilde{I}]-{T_l(\E[\zeta]-1)})\frac{C_{\textrm{sleep}}}{C_{\textrm{high}}}\right.
\nonumber +\frac{{\cal L}_{T_t}(\lambda)}{\lambda}
\frac{C_{\textrm{low}}}{C_{\textrm{high}}}\\
&&\left.+({T_l(\E[\zeta]-1)}+T_w){\cal L}_{T_t}(\lambda)
\frac{C_{\textrm{listen}}}{C_{\textrm{high}}}
\right)
\eear We expect the battery lifetime to
increase by the same factor.
In practice $C_{\textrm{sleep}}\ll C_{\textrm{high}}$ so that
terms in multiplication with $\frac{C_{\textrm{sleep}}}{C_{\textrm{high}}}$ can be
neglected. Letting $T_w=T_l$, the lifetime gain reduces to
\beq
\label{e:G}
G=\frac{{(1-\rho)\frac{C_{\textrm{low}}}{C_{\textrm{high}}}-\rho/\E[B]
\left({T_l\E[\zeta]}{\cal L}_{T_t}(\lambda)\frac{C_{\textrm{listen}}}{C_{\textrm{high}}}+\frac{{\cal L}_{T_t}(\lambda)}{\lambda}\frac{C_{\textrm{low}}}{C_{\textrm{high}}}\right)} }{\rho+(1-\rho)\frac{C_{\textrm{low}}}{C_{\textrm{high}}}}.
\eeq
All performance metrics found so far have been derived as
functions of
\begin{itemize}
\item {\em network} parameters: such as the load $\rho$, the input rate
$\lambda$, and the first and second moments of the service time ($\E[\sigma]$
and $\E[\sigma^2]$);
\item {\em physical} parameters: such as the consumption rates
$C_{\textrm{low}}$, $C_{\textrm{high}}$ and $C_{\textrm{listen}}$,
neglecting $C_{\textrm{sleep}}$;
\item {\em combined} physical and network parameters: such as the
listen window $T_l$ and warm-up period $T_w$;
\item the LSTs of the vacation periods and their first and second
moments.
\end{itemize}
In the following we will specify the distribution of the sleep
windows $\{S_i\}_i$ so as to compute explicitly $\{L_i(s)\}_i$, $\{\E[V_i]\}_i$
and $\{\E[V_i^2]\}_i$.
\subsection{Sleep Windows are Deterministic}
\label{s:S-const}
We will first consider that the sleep windows $\{S_i\}_i$ are deterministic. More precisely, let
\[
S_i=a^{\min\{i-1,l\} }T_{\min},\quad i=1,2,\ldots,
\]
where $T_{\min}$ is the initial sleep window size, $a$ is a multiplicative factor, and $l$ is the final sleep window exponent or equivalently the number of times the sleep window could be increased. We call $T_{\min}$, $a$ and $l$ the {\em protocol} parameters. The LSTs of the vacations periods and their first and second moments can be rewritten
\bears
{\cal L}_i(s)\;&=&\left\{
\barr{ll}
\exp(-T_{\min}s) , & i=1\\
&\\
\exp(-(a^{\min\{i-1,l\}}T_{\min}+T_l)s) , & i=2,3,\ldots,\\
\earr\right.\\
\E[V_i^n]&=&\left\{
\barr{ll}
T_{\min}^n , & i=1\\
&\\
(a^{\min\{i-1,l\}}T_{\min}+T_l)^n , & i=2,3,\ldots,\\
\earr\right.
\eears
for $n=1,2$.
We will study two cases so as to model type I and type
II saving classes as defined in the IEEE 802.16e standard (see
Sect.~\ref{s:intro}).
\subsubsection*{Scenario D-I}
This scenario is inspired by type I power saving classes \cite{man}. We consider
$a>1$ which implies that the first $l+1$ sleep windows are all
distinct. In particular, the value $a=2$ is consistent with IEEE
802.16e type I power saving classes.
\subsubsection*{Scenario D-II}
In order to mimic the type II power saving classes of the IEEE
802.16e, we set $a=1$ in this scenario. Letting $a=1$
equates the length of all sleep windows. Observe that we could have
alternatively let $l=0$; the resulting sleep windows would then be the
same, namely $S_i=T_{\min}$ for any $i$.
Recall from Sect.~\ref{s:intro} that in type II classes, a node may
send or receive traffic during listen windows if the requests
handling time is short enough. Hence, our model applies to these
classes only if we assume that no request is sufficiently small to be
served during a listen window $T_l$.
\subsection {Sleep Windows are Exponentially Distributed}
\label{s:S-exp}
As an alternative to deterministic sleep windows, we explore in this
section the situation when the sleep window $S_i$ is exponentially
distributed with parameter $\mu_i$, for $i=1,2,\ldots$. Similar to what was
done in Sect.~\ref{s:S-const}, we let
\beq
\E[S_i]=\frac{1}{\mu_i}=a^{\min\{i-1,l\} }T_{\min},\quad i=1,2,\ldots.
\label{e:Si-exp}
\eeq
The LSTs of the $\{V_i\}_i$ and their first and second moments are given
below.
\bears
{\cal L}_i(s)\,&=&\left\{
\barr{ll}
\displaystyle{\frac{1}{1+T_{\min}s}} , \quad i=1,\\
& \\
\displaystyle{\frac{\exp(-sT_l)}{1+a^{\min\{i-1,l\}}T_{\min}s}} , \quad i=2,3,\ldots,\\
\earr\right.\\
\E[V_i]\;&=&\left\{
\barr{ll}
T_{\min} , \quad i=1,\\
\\
a^{\min\{i-1,l\}}T_{\min}+T_l , \quad i=2,3,\ldots,\\
\earr\right.\\
\E[V_i^2]&=&\left\{
\barr{ll}
2T_{\min}^2 , \quad i=1\phantom{,3,\ldots.}\\
&\\
2a^{2\min\{i-1,l\}}T_{\min}^2+2a^{\min\{i-1,l\}}T_{\min}T_l+T_l^2 , \\
&\\\qquad\qquad\qquad\qquad\qquad i=2,3,\ldots.
\earr\right.
\eears
Like in Sect.~\ref{s:S-const}, we consider two cases inspired by the
first two types of IEEE 802.16e power saving classes.
\subsubsection*{Scenario E-I}
Similarly to what is considered in scenario D-I, we consider
multiplicative factors that are larger than 1, in other words, the
values $\{\mu_i\}_{i=1,\ldots,l+1}$ are different. When $a>1$, the sleep windows
increase in average over time. For $T_l=0$ we can find closed-form
expressions for all metrics derived in Sect.~\ref{s:ana}.
\subsubsection*{Scenario E-II}
The last case considered in this paper is when the sleep windows
are i.i.d. exponential random variables. This can be achieved by
letting either $a=1$ or $l=0$ in~\eqref{e:Si-exp}. Hence
$\mu_i=1/T_{\min}$ for any $i$. The LSTs of the $\{V_i\}_i$ and their
first and second moments simplify to
\bears
{\cal L}_i(s)\,&=&\left\{
\barr{ll}
\displaystyle{\frac{1}{1+T_{\min}s}}, \quad i=1,\\
&\\
\displaystyle{\frac{\exp(-sT_l)}{1+T_{\min}s}}\quad i=2,3,\ldots\\
\earr\right. \\
\E[V_i]\;&=&\left\{
\barr{ll}
T_{\min},\quad i=1,\\
&\\
T_{\min}+T_l,\quad i=2,3,\ldots\\
\earr\right.\\
\E[V_i^2]&=&\left\{
\barr{ll}
2T_{\min}^2, \quad i=1,\\
&\\
2T_{\min}^2+2T_{\min}T_l+T_l^2 ,\quad i=2,3,\ldots\\
\earr\right.
\eears
\section{Exploiting the Analytical Results}
\label{s:exploit}
Our model is useful for evaluating performance measures as a function of various network parameters (such as the input rate), and allows us to identify the protocol parameters that mostly impact the system performance. Instances of the expected system response time $T$ and the expected energy gain $G$ are provided in Sect.~\ref{s:ResPeva}.

Beside performance evaluation, we will use our analytical model to solve a large range of optimization problems. Below we propose some optimization problems adapted to various degrees of knowledge on the parameters defining the traffic statistics. 
\begin{enumerate}
\item {\bf Direct optimization:}
This approach is useful when the traffic parameters information (e.g. the arrival rate) are directly available, or when they can be measured or estimated. An optimization problem can thus be formulated to maximize the system performance (e.g. the energy gain); see Sect.~\ref{s:optim} for details.
\item
{\bf Average performance:} Given that we know the probability distribution of the traffic parameters then we may obtain the protocol parameters that optimize the expected system performance. This optimization analysis is detailed in Sect.~\ref{s:exp}.
\item
{\bf Worst case performance:} In the case where we do not have knowledge of even the statistical distribution of the network parameters, then we can formulate the worst case optimization problem which aims at guaranteeing the optimal performance under worst choice of network parameter. Though this is a more robust optimization approach, it yields a quite pessimistic selection of protocol parameters. Even if we do have knowledge of the statistical distribution, we may have to use a worst case performance in the case that that there is a strict bound on the value of some performance measure. The worst-case analysis will be further detailed in Sect.~\ref{s:worst}.
\end{enumerate}
We propose a multi objective formulation of the optimization problem, where the performance objectives are the energy consumption (or performance measures directly related to the energy consumption) and the response time. We formulate the multi objective problem as a constrained optimization one: the energy related criterion will be optimized under a constraint on the expected sojourn time. When the traffic parameters are not directly known, two types of constraints on the expected sojourn time will be considered; in the first case the constraint is with respect to the average performance, and in the second case, it is on the worst case performance.
\subsection{Constrained Optimization Problem}
\label{s:optim}
The objective is to optimize the protocol parameters defined earlier, namely, the initial window $T_{\min}$, the multiplicative factor $a$, and the exponent $l$. We define the following generic non-linear program:
\begin{subequations}
\label{e:opt}
\beq
\label{e:maxG}
\barr{ll}
\textrm{maximize}\;\; G , \quad
\textrm{subject to} \;\; T \leq T_{\textrm{QoS}} \\
\earr
\eeq
or equivalently (recall~\eqref{e:G-def})
\beq
\label{e:minE}
\barr{ll}
\textrm{minimize} \;\; E_{\textrm{sleep}} , \quad
\textrm{subject to}\;\; T \leq T_{\textrm{QoS}} \\
\earr
\eeq
\end{subequations}
where $G$ is given in~\eqref{e:G}, $E_{\textrm{sleep}}$ is given in~\eqref{e:Esleep1} and $T$, the system response time, is given in~\eqref{e:T-alt}. The program~\eqref{e:opt} maximizes the energy gain, or equivalently, minimizes the expected energy consumption rate, conditioned on a maximum system response time $T_{\textrm{QoS}}$. The value of $T_{\textrm{QoS}}$ is application-dependent; it needs to be small for interactive multimedia whereas larger values are acceptable for web traffic.

The decision variables in the above optimization will correspond to one or more protocol parameters. For a given distribution of the sleep windows $\{S_i\}_i$, the expected number of vacations $\E[\zeta]$, the expected idle period $\E[I]$, and subsequently the gain $G$ and the expected energy consumption rate $E_{\textrm{sleep}}$ will depend on the protocol parameters $T_{\min}$, $a$ and $l$ and on the physical parameters $C_{\textrm{low}}$, $C_{\textrm{high}}$ and $C_{\textrm{listen}}$ (assumed fixed).

We propose four types of applications of the mathematical program~\eqref{e:opt}.
\begin{enumerate}
\item In the first, denoted ${\cal P}_1$, the decision variable is the
initial expected sleep window $T_{\min}$. The parameters $a$ and $l$
are held fixed.
\item The second mathematical program, denoted ${\cal P}_2$, has as
decision variable the multiplicative factor $a$ whereas $T_{\min}$
and $l$ are given.
\item The decision variable of the third program, denoted ${\cal
P}_3$, is the exponent $l$. The parameters $T_{\min}$ and $a$ are
given.
\item In the fourth program, denoted ${\cal P}_4$, all three protocol
parameters are optimized. The corresponding energy gain $G$ is the
highest that can be achieved.
\end{enumerate}
These four mathematical programs will be solved considering (i)
deterministic and (ii) exponentially distributed sleep windows
$\{S_i\}_i$. Instances are provided in Sect.~\ref{s:ResOptim}.

\subsection{Expectation Analysis}
\label{s:exp}
Assume that the statistical distribution of the arrival process is known. Then we may obtain the protocol parameters that optimize the {\em expected} system performance. One may want to optimize either the
expected energy consumption in power save mode or the economy of energy achieved by activating the power save mode. These problems are not equivalent as was the case in~\eqref{e:opt} since the energy
consumption in normal mode itself also depends on the arrival process.

As already mentioned, we consider two different constraints on the expected sojourn time corresponding to the situations in which the application is sensitive either to the worst case value (hard constraint) or the average value (soft constraint). 
\subsubsection*{Hard Constraints}
Here, the application is very sensitive to the delay, so we need to ensure that the constraint on the expected sojourn time is always satisfied no matter the value of $\lambda$. The problem is to find the protocol parameter $\theta$ that achieves
\beq
\label{e:EH-E}
\barr{l}
\min_\theta \sum_\lambda p(\lambda)E_{\textrm{sleep}}(\lambda,\theta), \quad
\textrm{subject to } T(\lambda,\theta) \leq T_{\textrm{QoS}} \; \forall \,\,\lambda .\\
\earr
\eeq
Another problem is to find the protocol parameter $\theta$ that achieves
\beq
\label{e:EH-G}
\barr{l}
\max_\theta \sum_\lambda p(\lambda)G(\lambda,\theta) , \quad
\textrm{subject to } T(\lambda,\theta) \leq T_{\textrm{QoS}} \; \forall\,\, \lambda .\\
\earr
\eeq
The problems~\eqref{e:EH-E} and~\eqref{e:EH-G} are not equivalent
because $G$ depends also on $E_{\textrm{no sleep}}$ which itself
depends on $\lambda$; recall~\eqref{e:Enosleep}. Instances of~\eqref{e:EH-G}
will be provided in Sect.~\ref{s:WorstExp}.
\subsubsection*{Soft Constraints}
In this optimization problem it is assumed that the application is sensitive only to the expected sojourn time rather than to its worst case value. The objective is to find $\theta$ that achieves
\beq
\label{e:ES-E}
\barr{l}
\min_\theta \sum_\lambda p(\lambda)E_{\textrm{sleep}}(\lambda,\theta) , \quad
\textrm{subject to } \sum_\lambda p(\lambda) T(\lambda,\theta) \leq T_{\textrm{QoS}} .\\
\earr
\eeq
Alternatively, one may want to find $\theta$ that achieves
\beq
\label{e:ES-G}
\barr{l}
\max\theta \sum_\lambda p(\lambda)G(\lambda,\theta) , \quad
\textrm{subject to } \sum_\lambda p(\lambda) T(\lambda,\theta) \leq T_{\textrm{QoS}} .\\
\earr
\eeq
Instances of~\eqref{e:ES-G} will be provided in
Sect.~\ref{s:WorstExp}.
\subsection{Worst Case Analysis}
\label{s:worst}
When the actual input rate is unknown, then a worst case analysis can be performed to enhance the performance under the considered time constraint. Let $\theta$ represent the protocol parameter(s) over which we optimize.
\subsubsection*{Hard Constraints}
Assume the constraint on the expected sojourn time has to be satisfied for any value of $\lambda$. The problem then is to find $\theta$ that achieves
\beq
\label{e:WH-E}
\barr{l}
\min_\theta \max_\lambda E_{\textrm{sleep}}(\lambda,\theta), \quad
\textrm{subject to } T(\lambda,\theta) \leq T_{\textrm{QoS}} \; \forall \,\,\lambda.\\
\earr
\eeq
In other words, we want to find the value of $\theta$ that improves the worst possible energy consumption.
A different problem consists of finding $\theta$ that improves the worst possible gain, namely,
\beq
\label{e:WH-G}
\barr{l}
\max_\theta \min_\lambda G(\lambda,\theta), \quad
\textrm{subject to } T(\lambda,\theta) \leq T_{\textrm{QoS}} \; \forall \,\,\lambda.\\
\earr
\eeq
Observe that the worst possible gain is the one obtained when the traffic input rate tends to $1/\E[\sigma]$. Thus $\min_\lambda G(\lambda,\theta)\approx0$. Therefore, the above problem is meaningful only for a restricted range of small values of $\lambda$ for which the worst energy gain is far above 0. Instances of~\eqref{e:WH-G} will be provided in Sect.~\ref{s:WorstExp}.
\subsubsection*{Soft Constraints}
Here, the application is not very sensitive to the delay, so it is acceptable that the constraint is respected by the average performance. The statistical distribution of the input rate, denoted $p(\lambda)$, is assumed to be known. The problem is to find $\theta$ that achieves
\beq
\label{e:WS-E}
\barr{l}
\min_\theta \max_\lambda E_{\textrm{sleep}}(\lambda,\theta), \quad
\textrm{subject to } \sum_\lambda p(\lambda) T(\lambda,\theta) \leq T_{\textrm{QoS}} .\\
\earr
\eeq
Again, a different objective can be desired, namely to maximize the worst gain. Like what was mentioned in the previous section, the problem is meaningful only when the rate $\lambda$ is small.
\beq
\label{e:WS-G}
\barr{l}
\max_\theta \min_\lambda G(\lambda,\theta), \quad
\textrm{subject to } \sum_\lambda p(\lambda) T(\lambda,\theta) \leq T_{\textrm{QoS}} .\\
\earr
\eeq
Instances of~\eqref{e:WS-G} will be provided in Sect.~\ref{s:WorstExp}.

\section{Results and Discussion}
\label{s:res}
We have performed an extensive numerical analysis to evaluate the performance of the system in terms of the expected system response time $T$ given in~\eqref{e:T-alt} and the expected energy gain $G$ given in~\eqref{e:G}; cf. Sect.~\ref{s:ResPeva}. In addition we have solved the problems ${\cal P}_1$--${\cal P}_4$ for given values of the protocol parameters held fixed; cf. Sect.~\ref{s:ResOptim}. Instances of the problems~\eqref{e:EH-G}, \eqref{e:ES-G}, \eqref{e:WH-G} and \eqref{e:WS-G} are also provided; cf. Sect.~\ref{s:WorstExp}. We first consider vacation trigger time $T_t=0$, i.e. whenever the system is idle it is bound to go for at least one vacation. Latter we illustrate the impact of vacation trigger time depicting one of the cases.

Physical and network parameters have been selected as follows:
\[
\barr{lll}
{C_{\textrm{low}}}/{C_{\textrm{high}}}={ C_{\textrm{listen}}}/{C_{\textrm{high}}}=0.2 , \quad \E[\sigma]= 1 ,\quad \E[\sigma^2]=2 , \\
T_l= T_w=1,\quad T_{\textrm{QoS}}=50/100.
\earr
\]
Unless otherwise specified, the protocol parameters are set to the
{\em default} values: $T_{\min}=2$, $a=2$, $l=9$ and $T_t=0$ in scenarios D-I and
E-I, and $T_{\min}=2$, $a=1$, $l=0$ and $T_t=0$ in scenarios D-II and E-II.

We have varied $\lambda$ in the interval $(0,1)$, $T_{\min}$ in $(1,100)$, $a$
in $(1,10)$, and $T_t$ in $(0,10)$. The parameter $l$ takes integer values in the interval $(0,10)$.
\subsection{Performance Evaluation}
\label{s:ResPeva}
We have evaluated numerically the expected sojourn time $T$ and the expected energy gain $G$ in all four scenarios defined in Sects.~\ref{s:S-const} and~\ref{s:S-exp}, varying the input rate $\lambda$ and the three protocol parameters $T_{\min}$, $a$ and $l$. Our results will be presented in the following sections. First, we discuss the impact of each of the three parameters on the performance of the system in terms of $T$ and $G$: impact of $T_{\min}$ in Sect.~\ref{s:Tmin}, impact of $a$ in Sect.~\ref{s:a}, and impact of
$l$ in Sect.~\ref{s:l}. Then, we comment on each of the performance metrics: comments on $T$ are in Sect.~\ref{s:T}, and comments on $G$ are in Sect.~\ref{s:G}.
\subsubsection{Impact of the initial window size $T_{\min}$}
\label{s:Tmin}
\begin{figure*}[tb]
\begin{center}
\subfigure[sojourn time, deterministic sleep windows]{
\label{f:Tmin-T-DI}
\epsfig{file=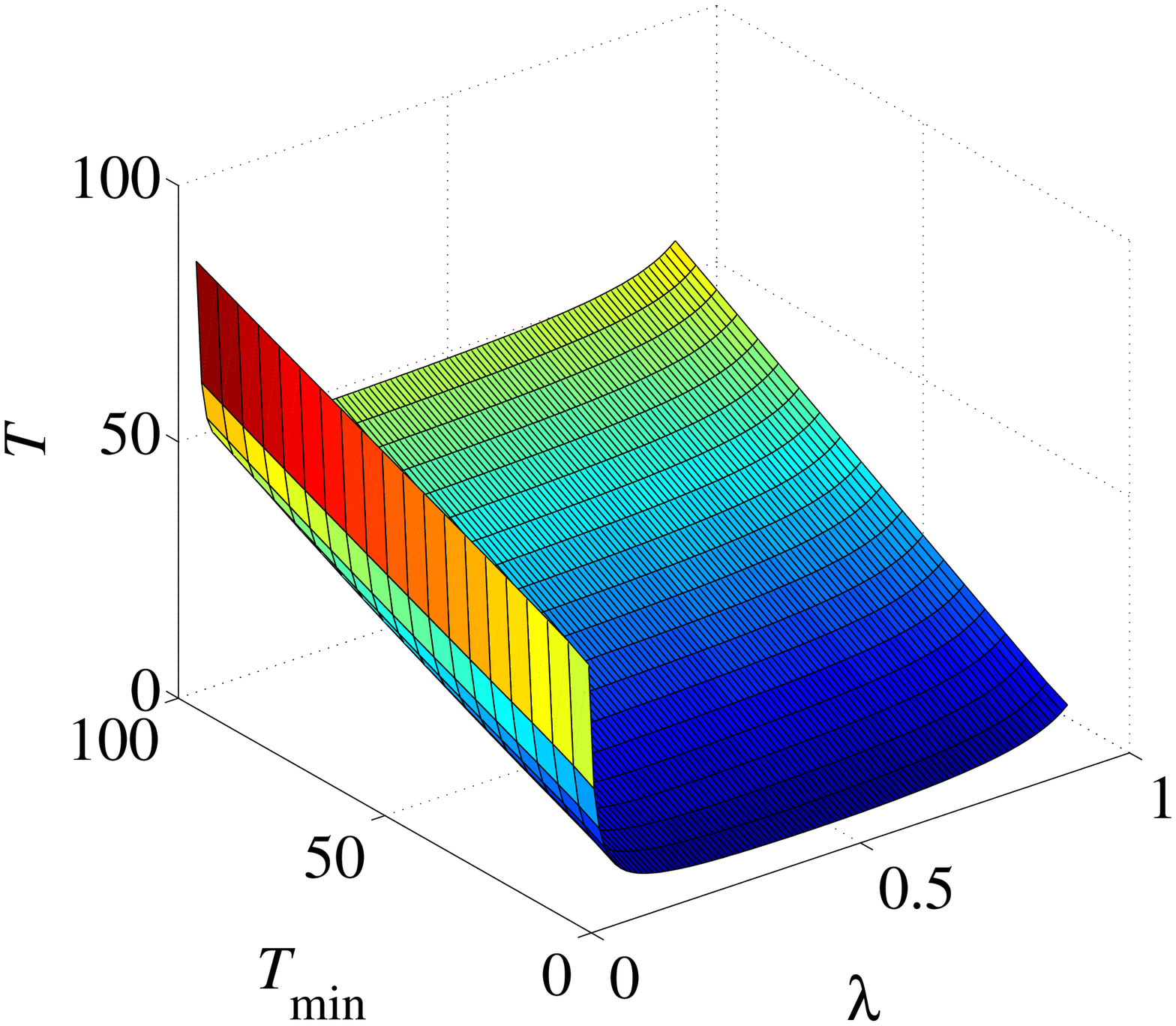,width=9pc}}
\subfigure[sojourn time, exponential sleep windows]{
\label{f:Tmin-T-EI}
\epsfig{file=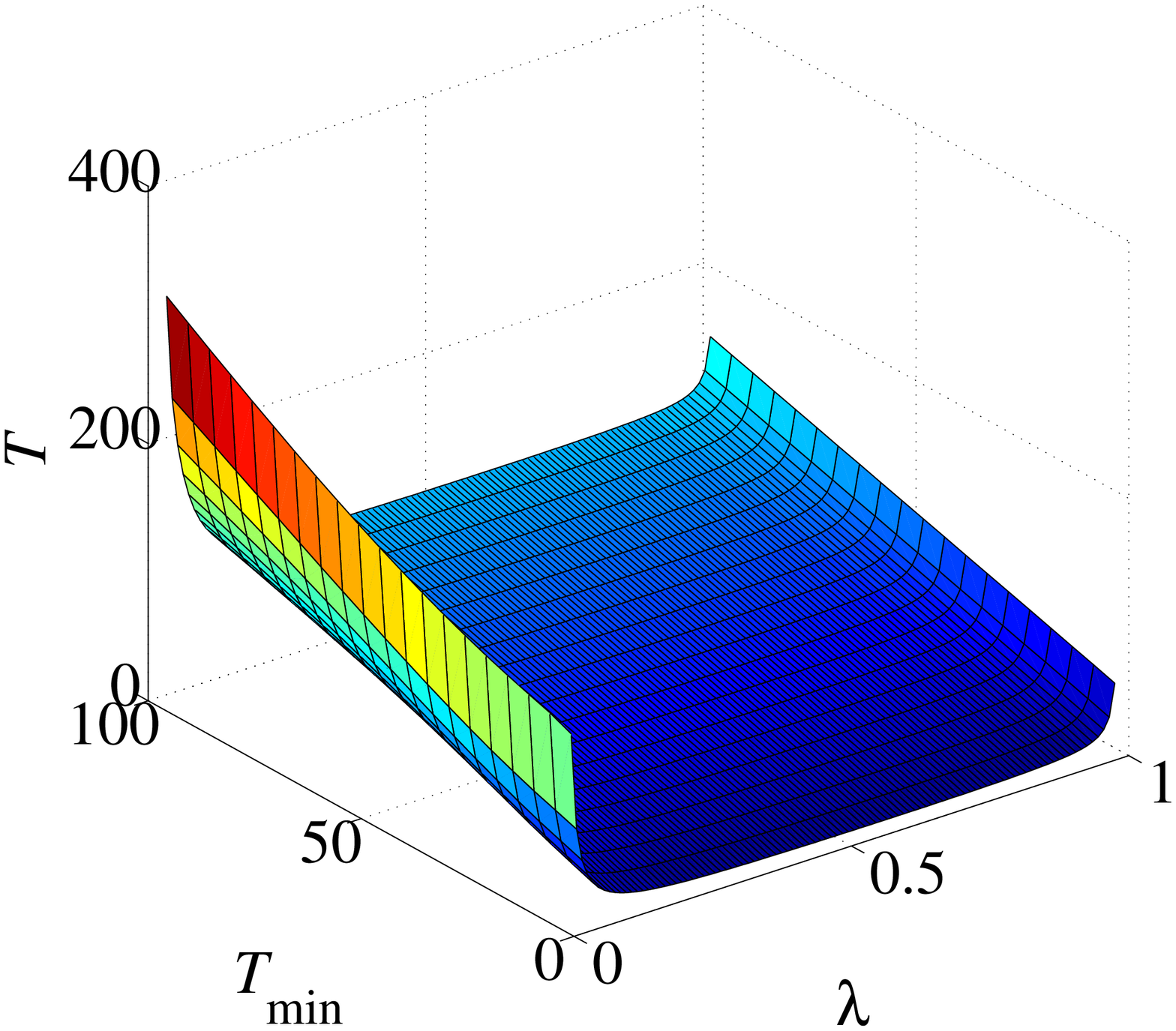,width=9pc}}
\subfigure[energy gain, deterministic sleep windows]{
\label{f:Tmin-G-DI}
\epsfig{file=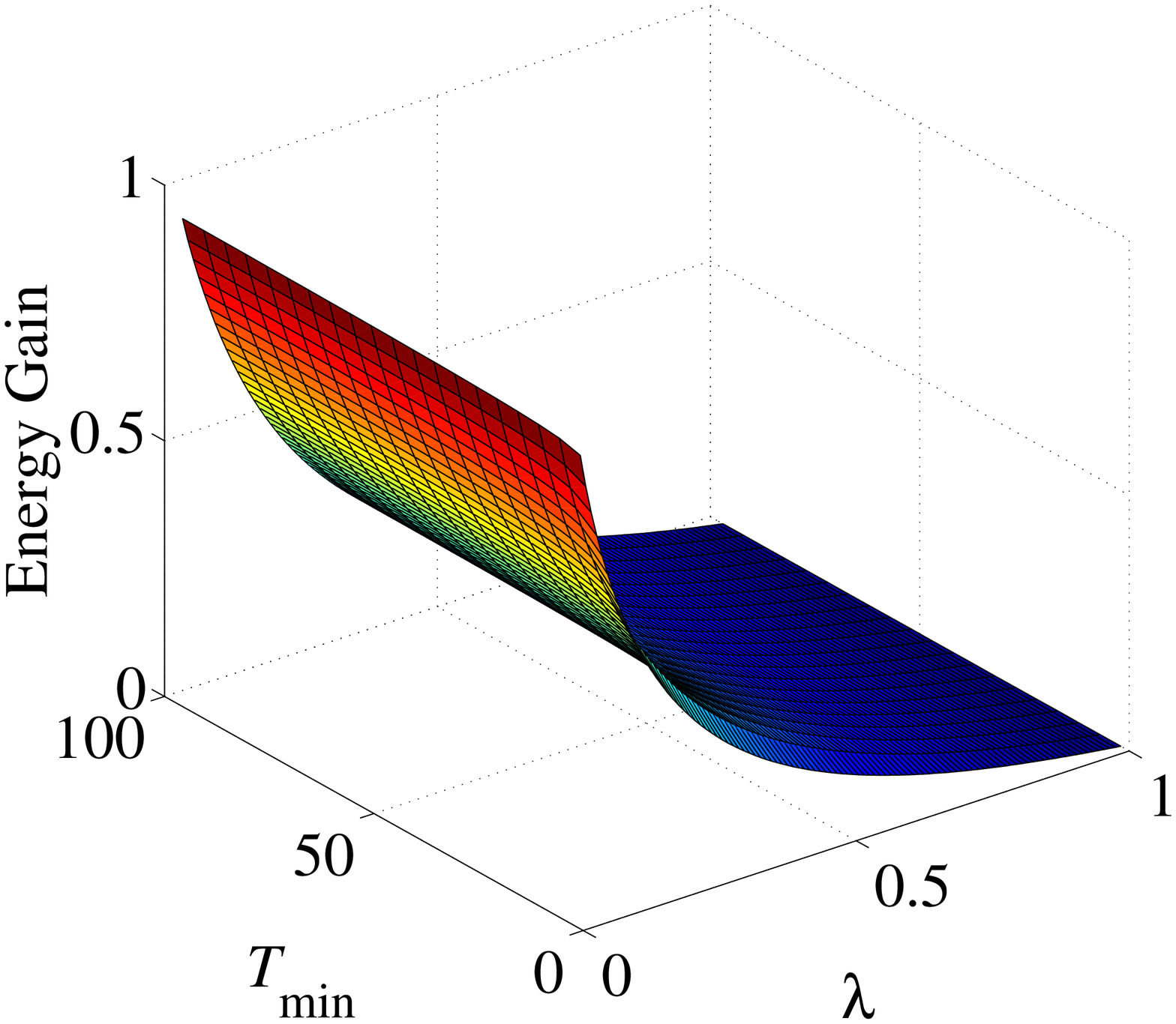,width=9pc}}
\subfigure[energy gain, exponential sleep windows]{
\label{f:Tmin-G-EI}
\epsfig{file=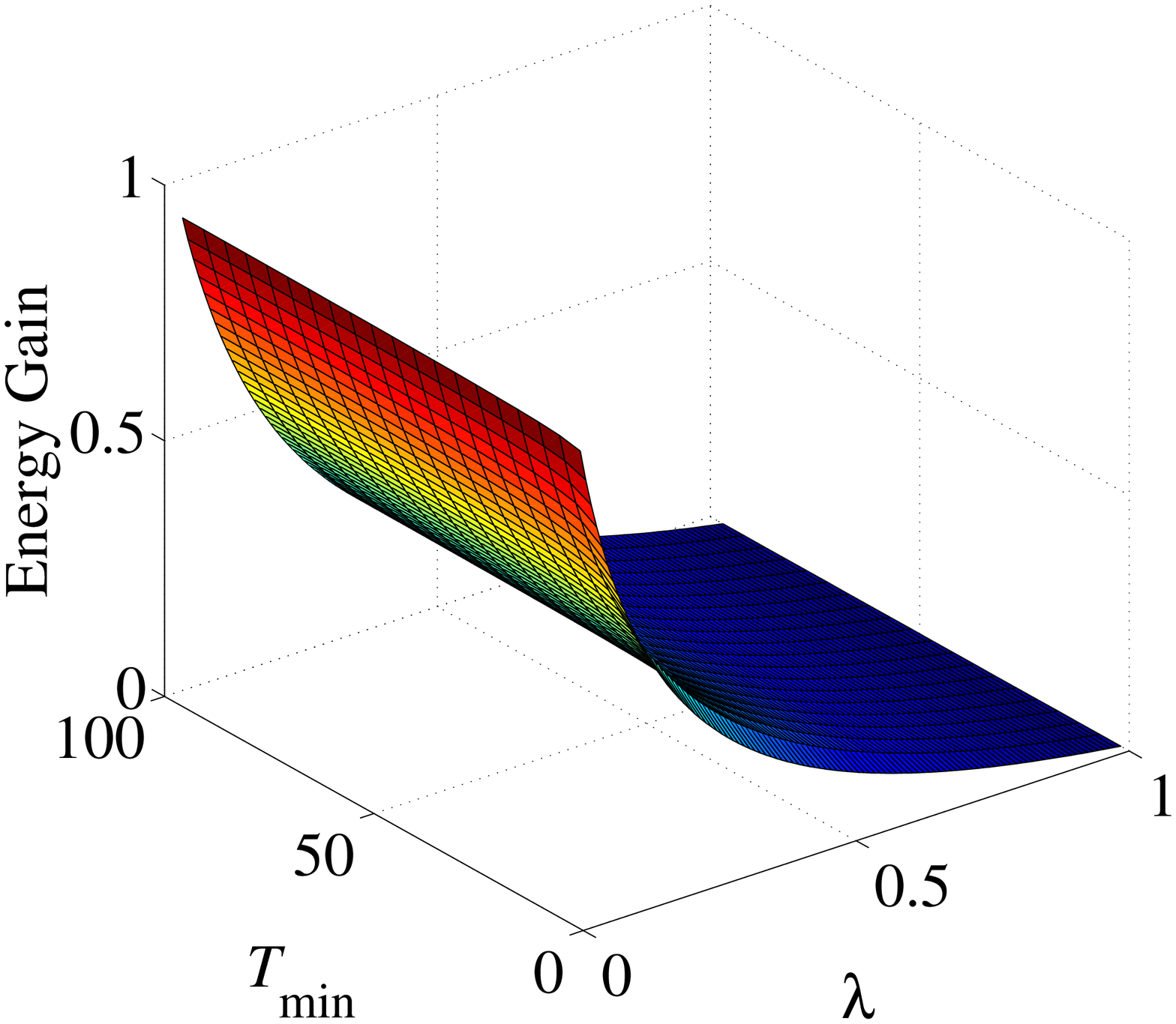,width=9pc}}
\caption{Impact of $T_{\min}$ on $T$ and $G$ in type I like power saving classes.
\label{f:Tmin-I}}
\end{center}
\end{figure*}

We will first investigate the impact that the initial window size $T_{\min}$ has on the performance of the system. For reasons that will be made clear later, this parameter is foreseen to be the most important parameter in type I like power saving classes (scenarios D-I and E-I) and it is the unique parameter in type II like power saving classes (scenarios D-II and E-II).
\paragraph{Type I like power saving classes}
We set $a=2$, $T_t=0$ and $l=9$ in scenarios D-I and E-I. The results are
graphically reported in Fig.~\ref{f:Tmin-I}.
Figures~\ref{f:Tmin-T-DI} and~\ref{f:Tmin-T-EI} respectively depict the expected sojourn time $T$ against the traffic input rate $\lambda$ and the initial sleep window size $T_{\min}$ when sleep windows are deterministic and exponentially distributed. The energy gain under the same conditions is depicted in Figs.~\ref{f:Tmin-G-DI} and~\ref{f:Tmin-G-EI}.

\begin{figure*}[tb]
\begin{center}
\subfigure[sojourn time, deterministic sleep windows]{
\label{f:T-DII}
\epsfig{file=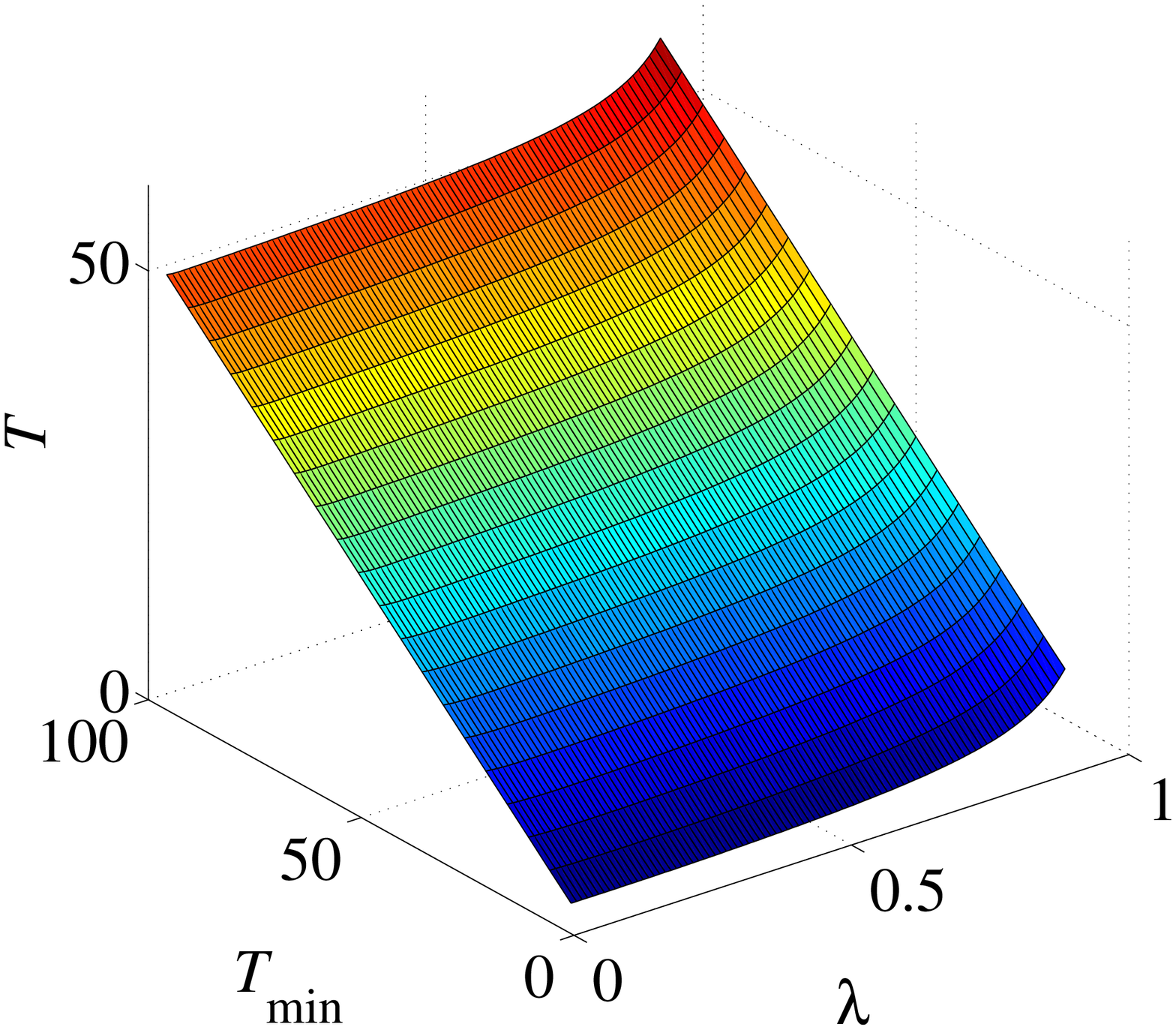,width=9pc}}
\subfigure[sojourn time, exponential sleep windows]{
\label{f:T-EII}
\epsfig{file=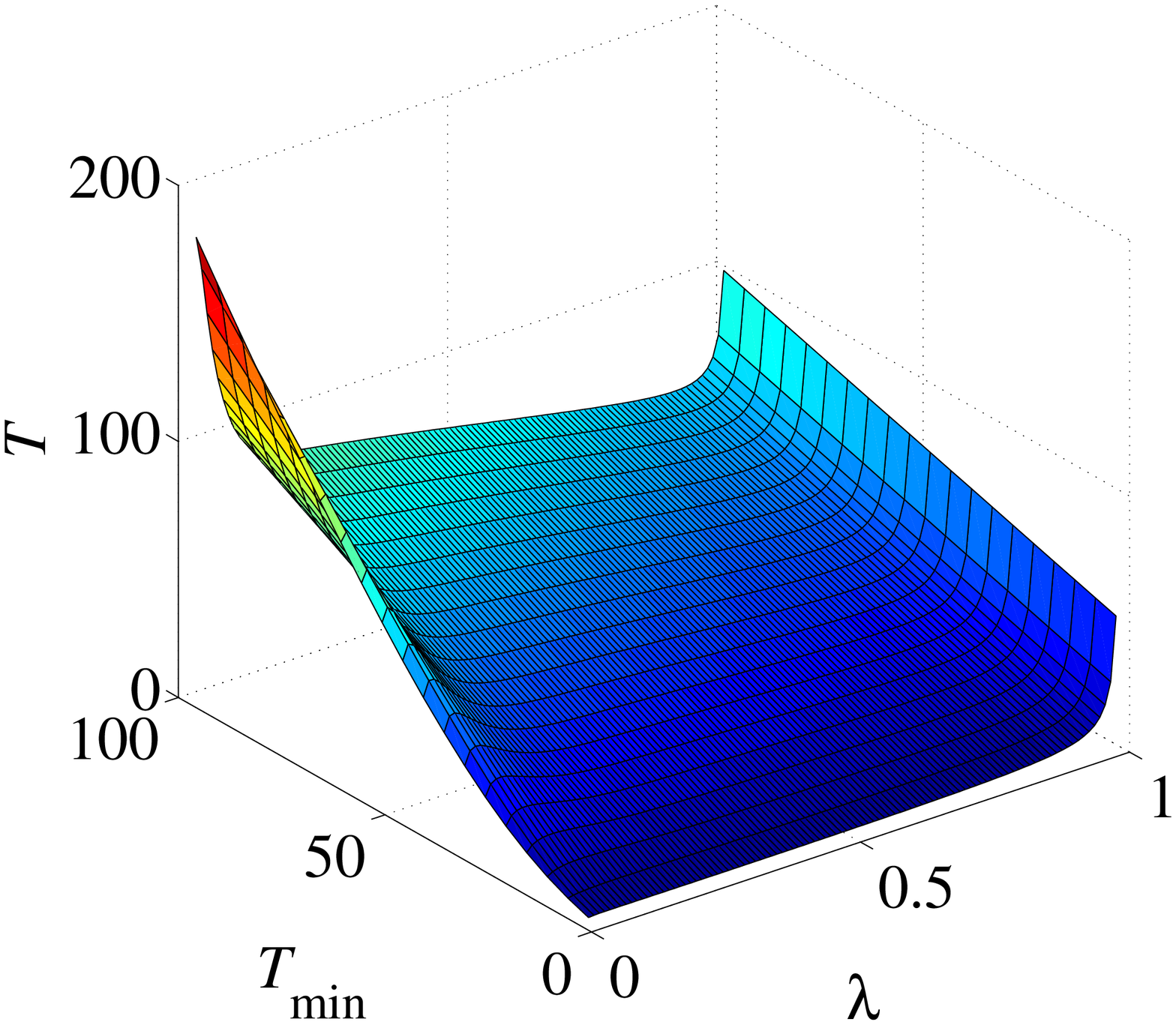,width=9pc}}
\subfigure[energy gain, deterministic sleep windows]{
\label{f:G-DII}
\epsfig{file=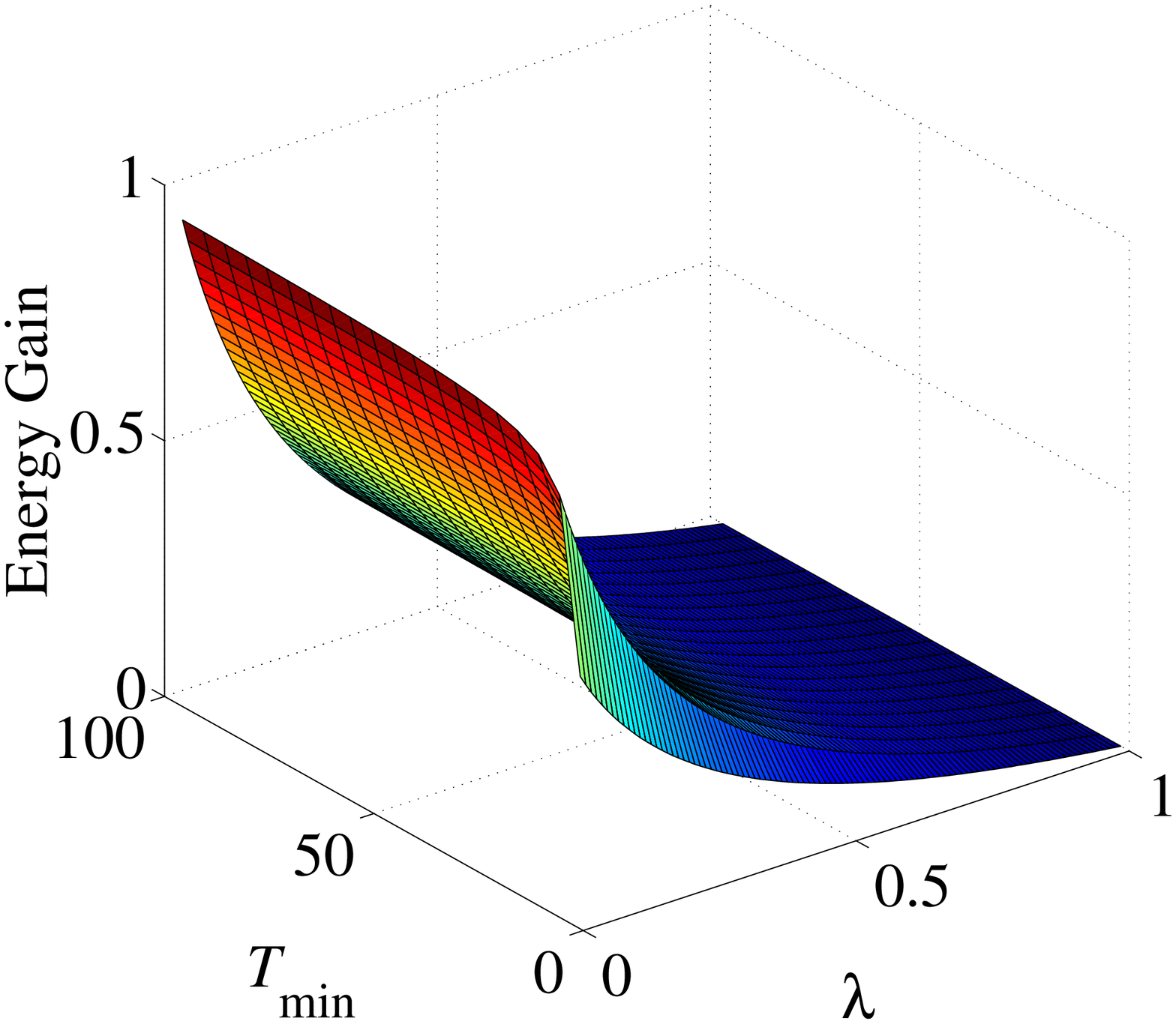,width=9pc}}
\subfigure[energy gain, exponential sleep windows]{
\label{f:G-EII}
\epsfig{file=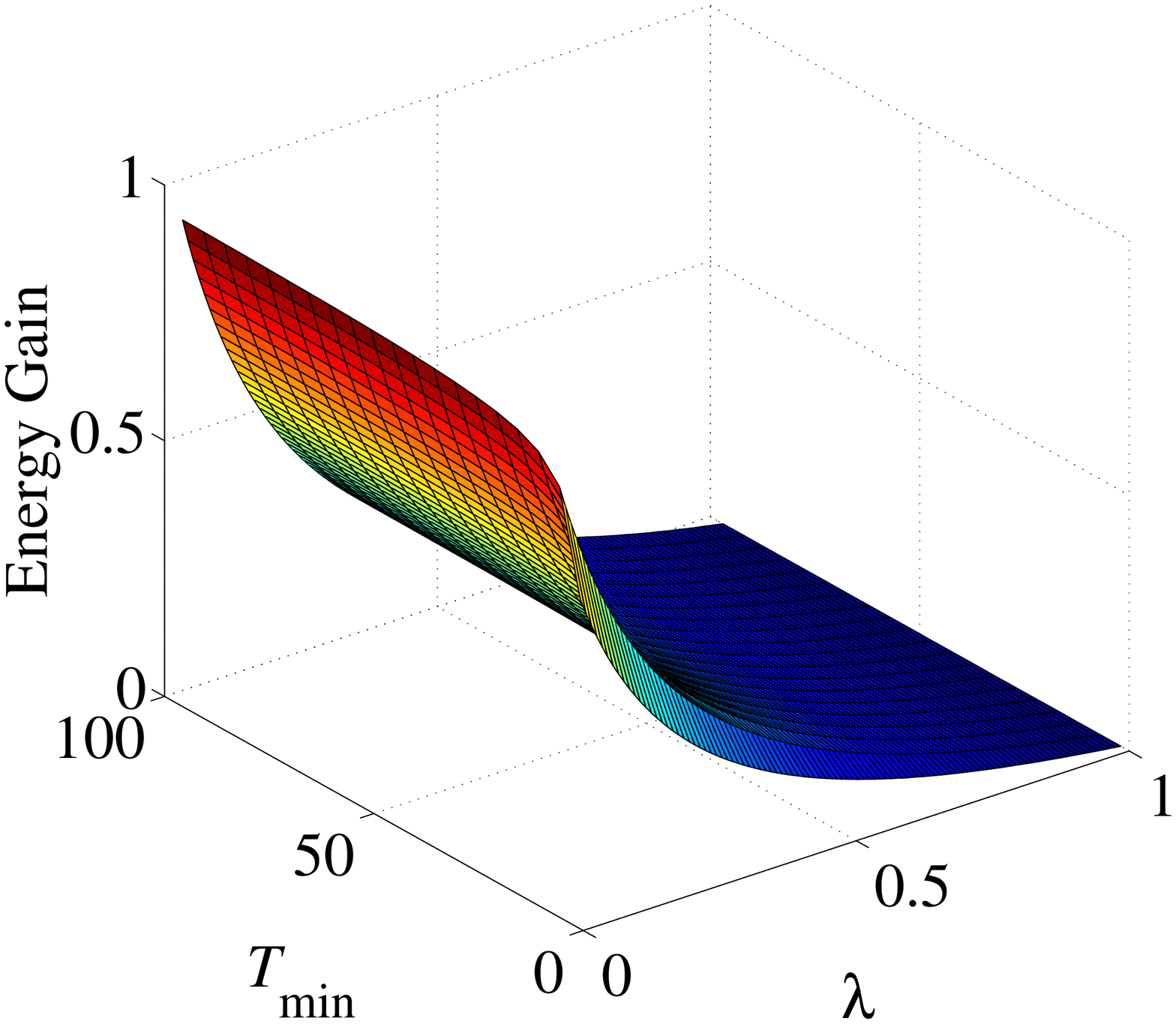,width=9pc}}
\caption{Impact of $T_{\min}$ on $T$ and $G$ in type II like power saving classes.
\label{f:Tmin-II}}
\end{center}
\end{figure*}

The size of the initial sleep window has a large impact on $T$ for any value of $\lambda$. More precisely, $T$ increases linearly with an increasing $T_{\min}$ for any $\lambda$; see Figs.~\ref{f:Tmin-T-DI},
\ref{f:Tmin-T-EI}. As for the gain $G$, it is not impacted by $T_{\min}$, except for a small degradation at very small values of $T_{\min}$, hardly visible in Figs.~\ref{f:Tmin-G-DI} and~\ref{f:Tmin-G-EI}.
\paragraph{Type II like power saving classes}
We set $a=1$, $T_t=0$ and $l=0$ in scenarios D-II and E-II. The results are graphically
reported in Fig.~\ref{f:Tmin-II}.
Figures~\ref{f:T-DII} and~\ref{f:T-EII} respectively depict the expected sojourn time $T$ against the traffic input rate $\lambda$ and the initial sleep window size $T_{\min}$ when sleep windows are
deterministic and exponentially distributed. The energy gain under the same conditions is depicted in Figs.~\ref{f:G-DII} and~\ref{f:G-EII}.

About the impact of $T_{\min}$ on $T$ and $G$, we can make similar observations to those made for type I like power saving classes, to the only exception that here the degradation of $G$ at very small values of $T_{\min}$ is more visible, especially in Fig.~\ref{f:G-DII}. 

Observe that a larger $T_{\min}$ yields a larger sleep time but it also reduces $\E[\zeta]$ which together explains why the impact on the energy gain is not significant.
\subsubsection{Impact of the multiplicative factor $a$}
\label{s:a}
\begin{figure*}[tb]
\begin{center}
\subfigure[sojourn time, deterministic sleep windows]{
\label{f:a-T-DI}
\epsfig{file=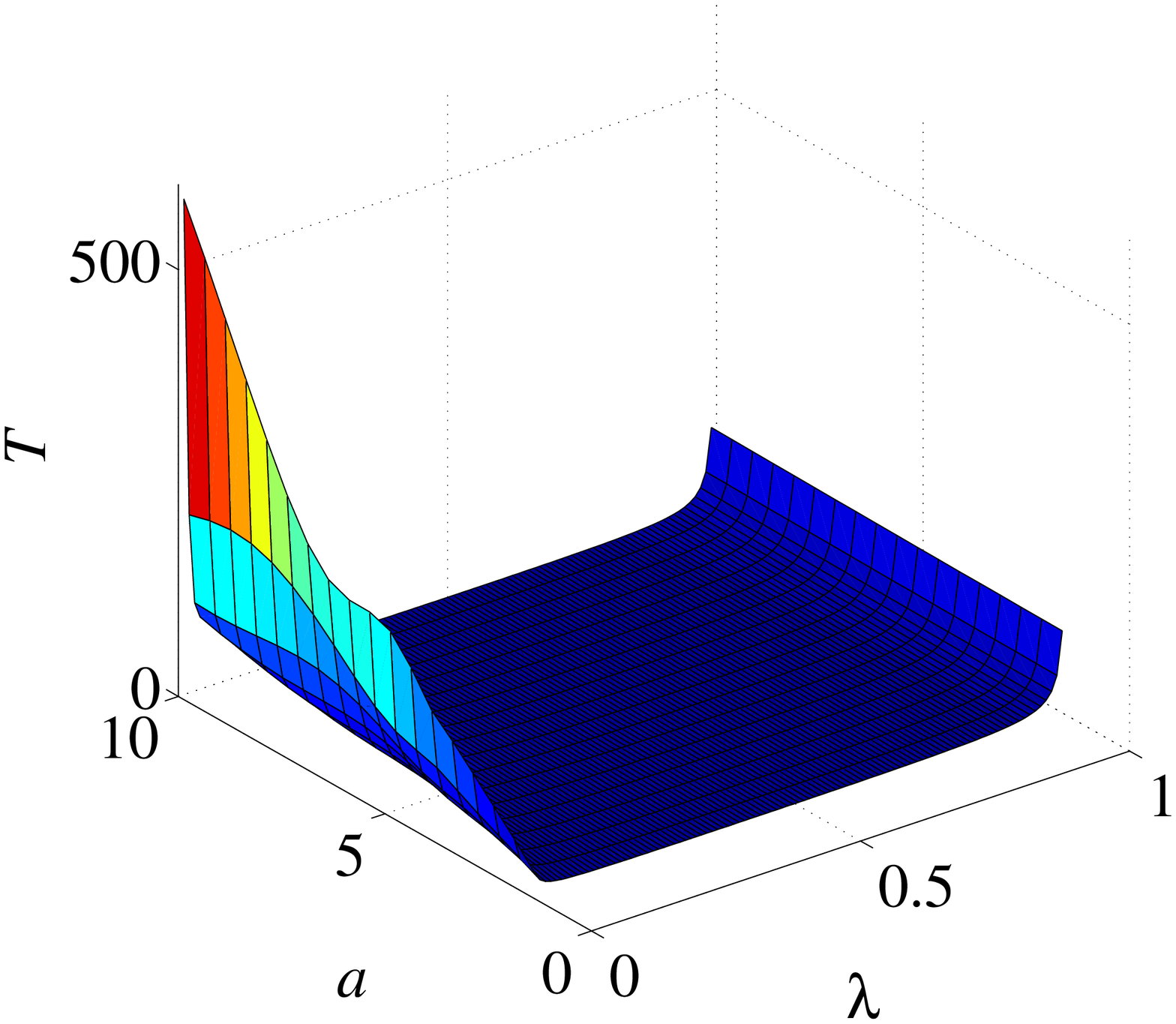,width=9pc}}
\subfigure[sojourn time, exponential sleep windows]{
\label{f:a-T-EI}
\epsfig{file=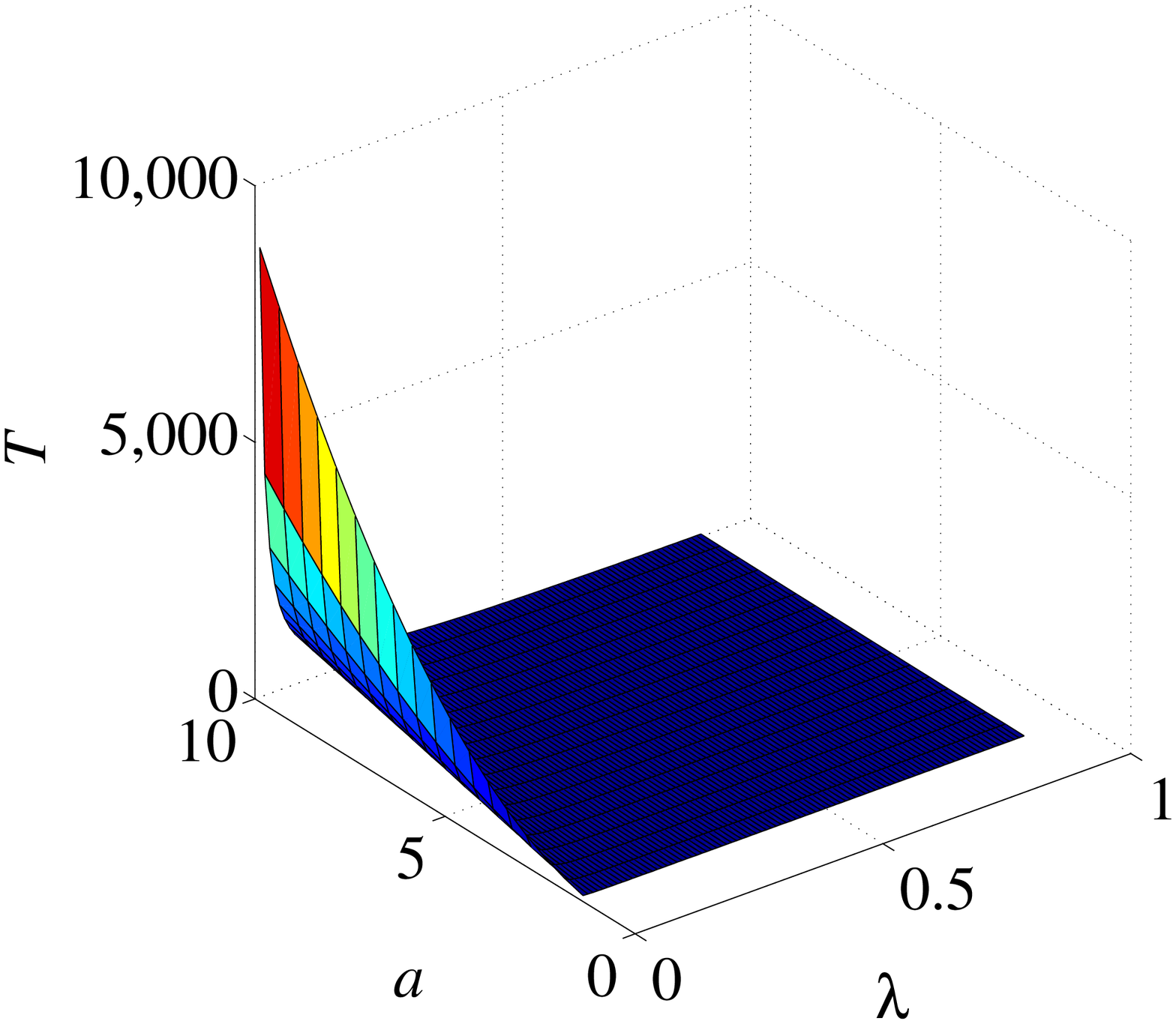,width=9pc}}
\subfigure[energy gain, deterministic sleep windows]{
\label{f:a-G-DI}
\epsfig{file=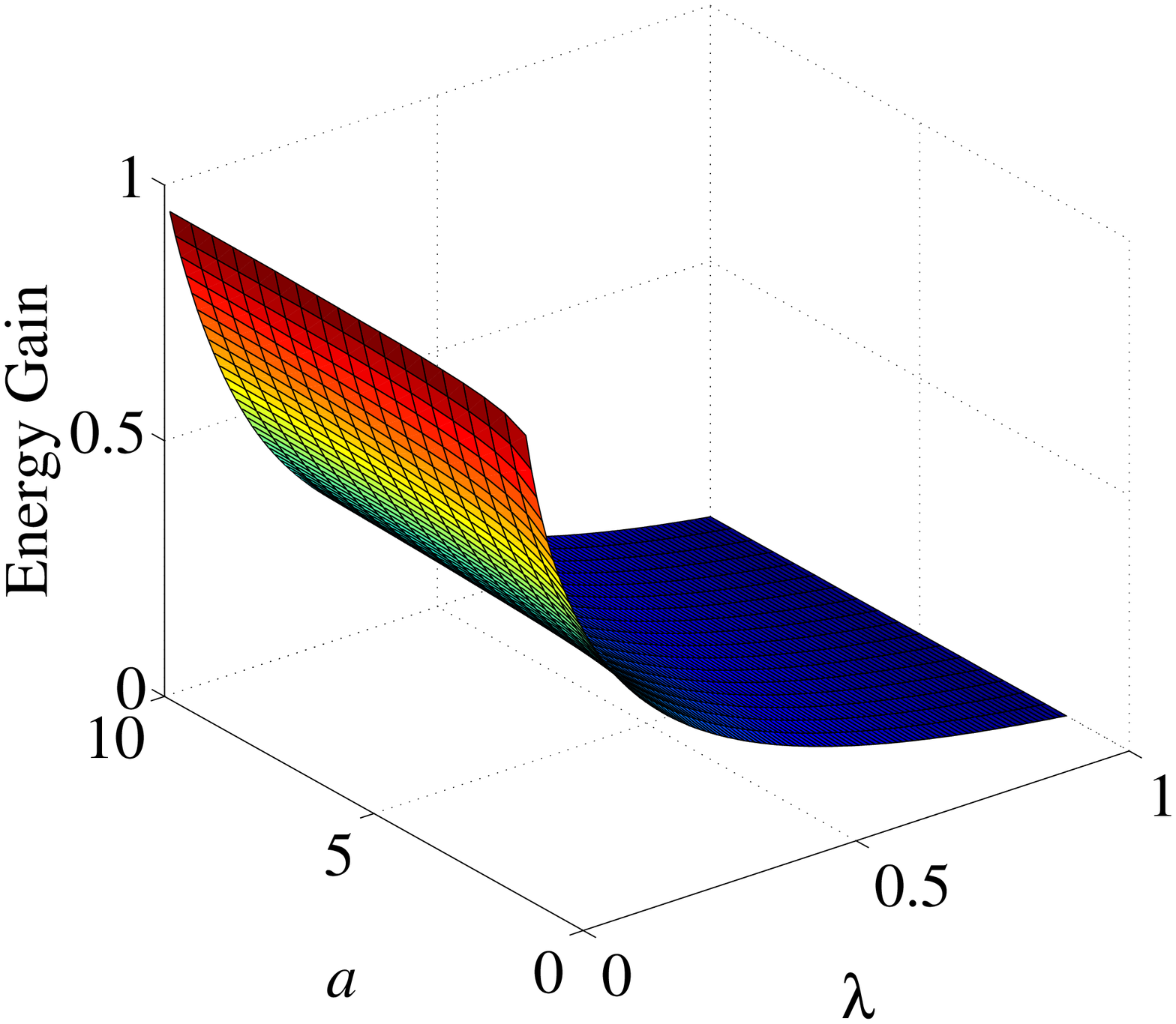,width=9pc}}
\subfigure[energy gain, exponential sleep windows]{
\label{f:a-G-EI}
\epsfig{file=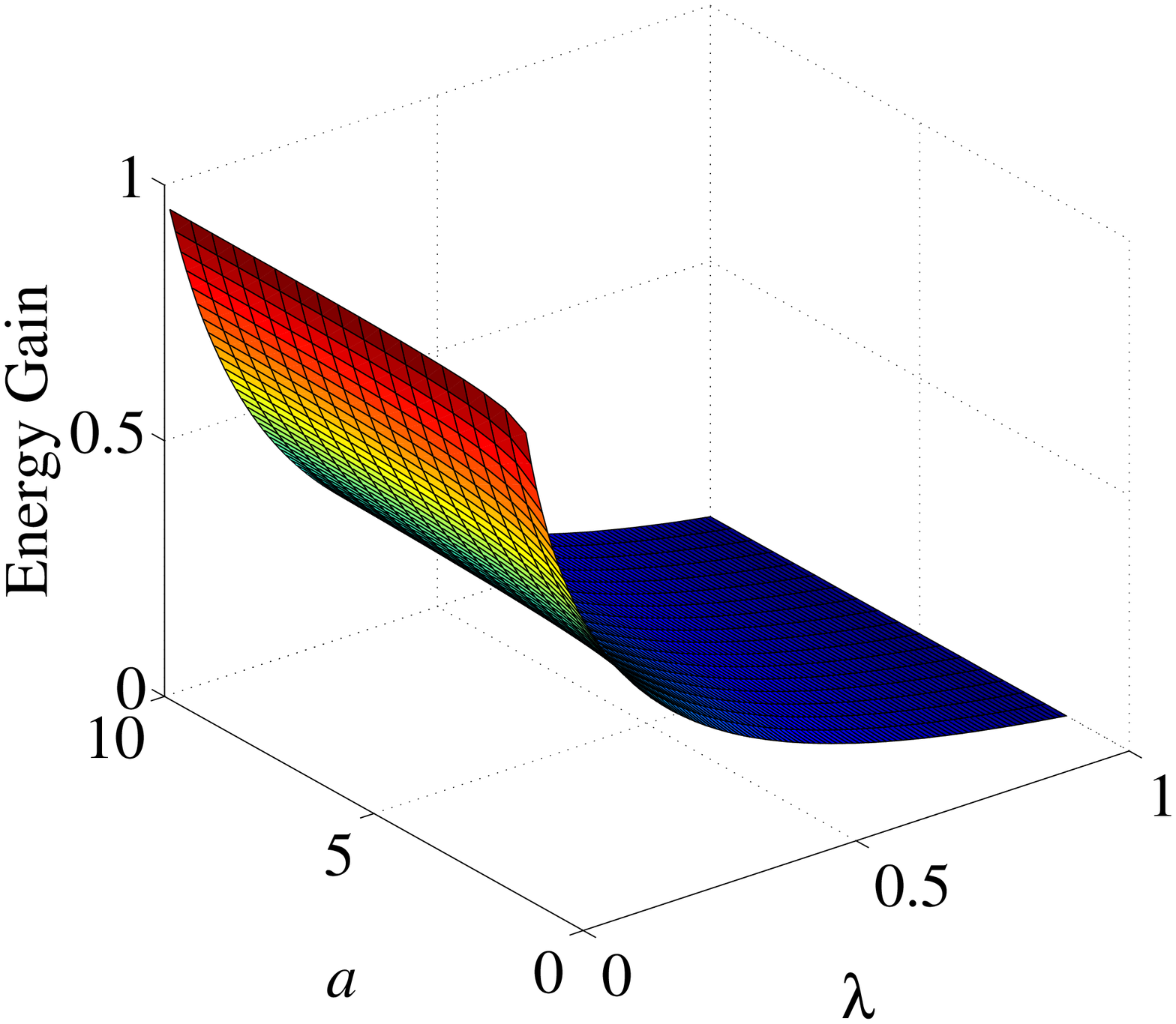,width=9pc}}
\caption{Impact of $a$ on $T$ and $G$ with either deterministic or exponential $\{S_i\}_i$.
\label{f:a-I}}
\end{center}
\end{figure*}

The second parameter used in type I like power saving classes (scenarios D-I and E-I) is the multiplicative factor $a$. In order to assess the impact of $a$ on the performance of the system, we perform a numerical analysis in which the initial window size is $T_{\min}=2$, the vacation trigger time $T_t=0$, the exponent is $l=9$ and the multiplicative factor $a$ is varied from 1 to 10. We evaluate the expected sojourn time $T$ and the energy gain $G$ both for deterministic (scenario D-I) and exponentially distributed (scenario E-I) sleep windows. We report the results inFig.~\ref{f:a-I}.
Figures~\ref{f:a-T-DI} and~\ref{f:a-G-DI} respectively depict the expected sojourn time $T$ and the energy gain $G$ against the traffic input rate $\lambda$ and the multiplicative factor $a$ when sleep windows are deterministic. The results obtained when the sleep windows are exponentially distributed are displayed in Figs.~\ref{f:a-T-EI} and~\ref{f:a-G-EI}. 

Interestingly enough, the multiplicative factor $a$ does not impact the gain $G$. It impacts greatly $T$ but only at very low input rates. Observe that $T$ increases exponentially with an increasing $a$ for small
$\lambda$ which is reflected in Figs.~\ref{f:a-T-DI} and \ref{f:a-T-EI}. 

\subsubsection{Impact of the exponent $l$} \label{s:l}
\begin{figure*}[tb]
\begin{center}
\subfigure[sojourn time, deterministic sleep windows]{
\label{f:l-T-DI}
\epsfig{file=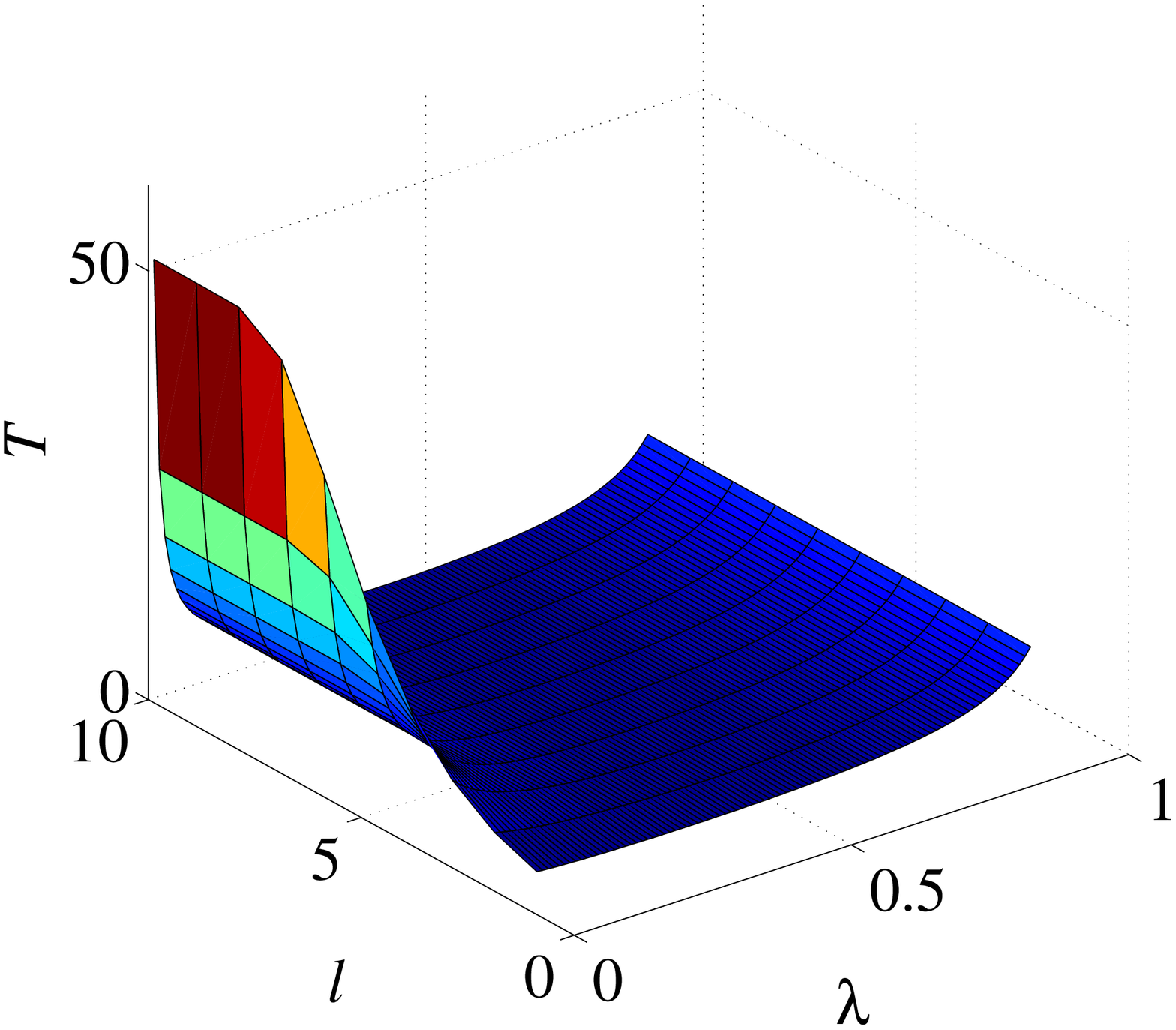,width=9pc}}
\subfigure[sojourn time, exponential sleep windows]{
\label{f:l-T-EI}
\epsfig{file=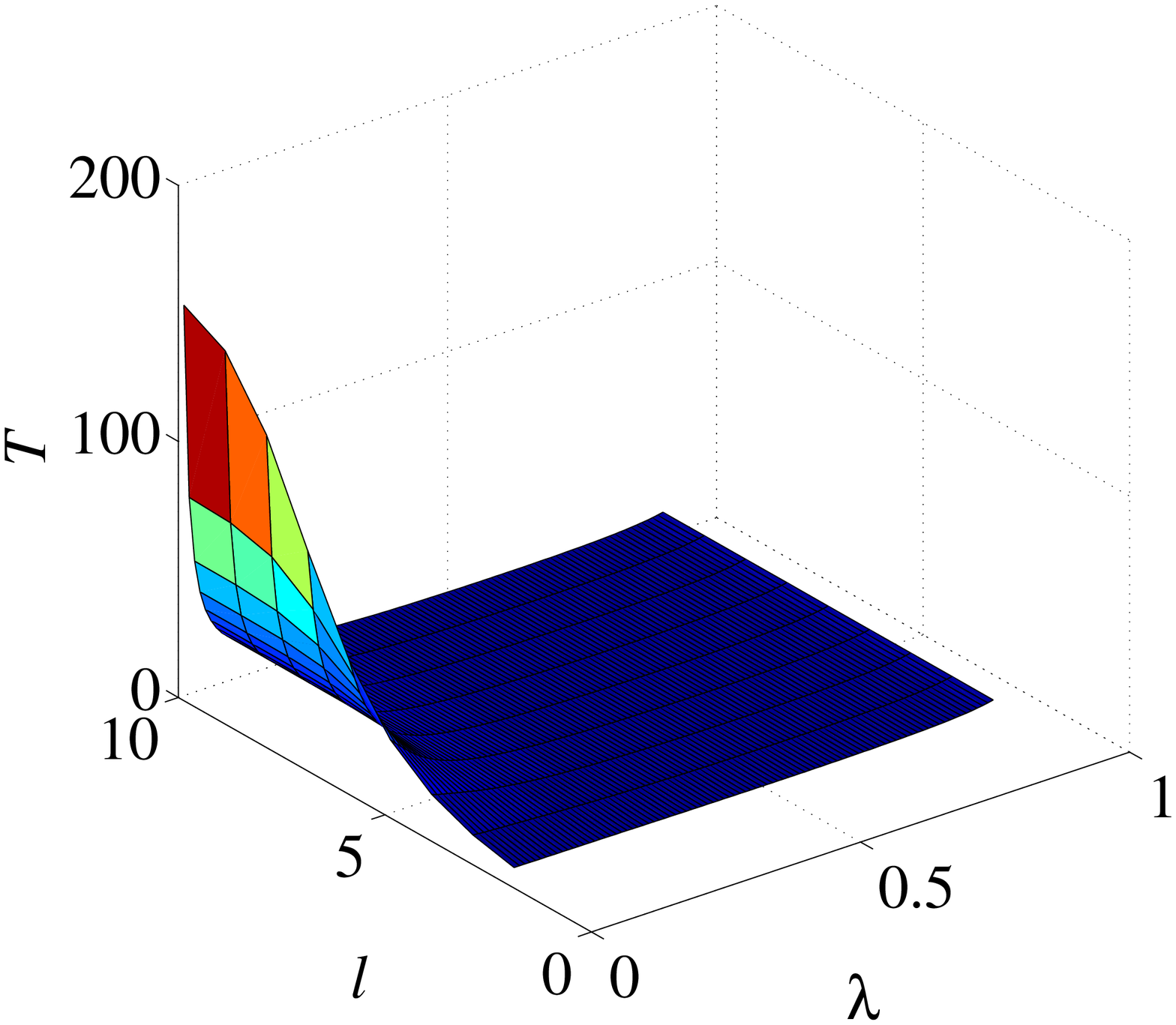,width=9pc}}
\subfigure[energy gain, deterministic sleep windows]{
\label{f:l-G-DI}
\epsfig{file=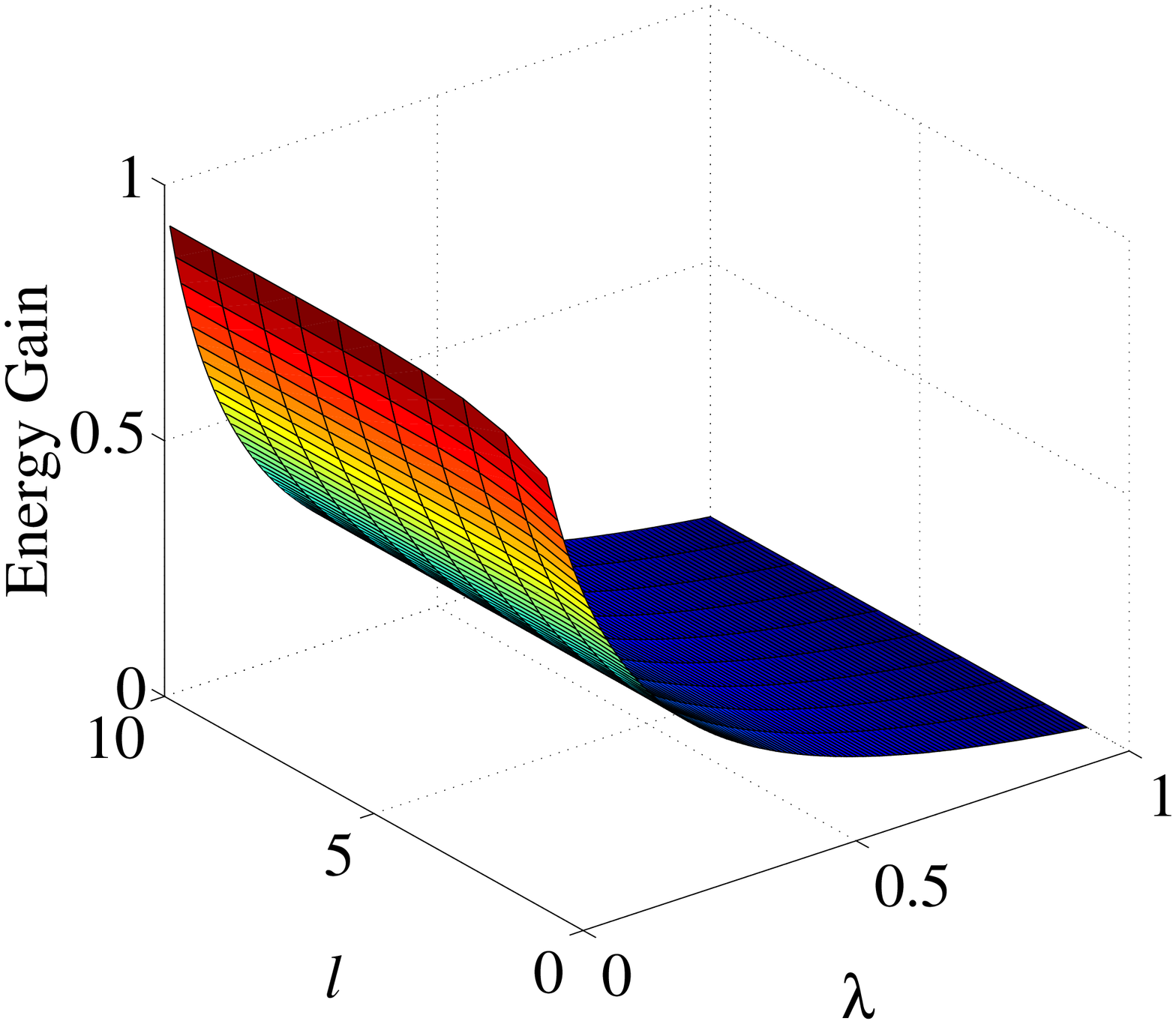,width=9pc}}
\subfigure[energy gain, exponential sleep windows]{
\label{f:l-G-EI}
\epsfig{file=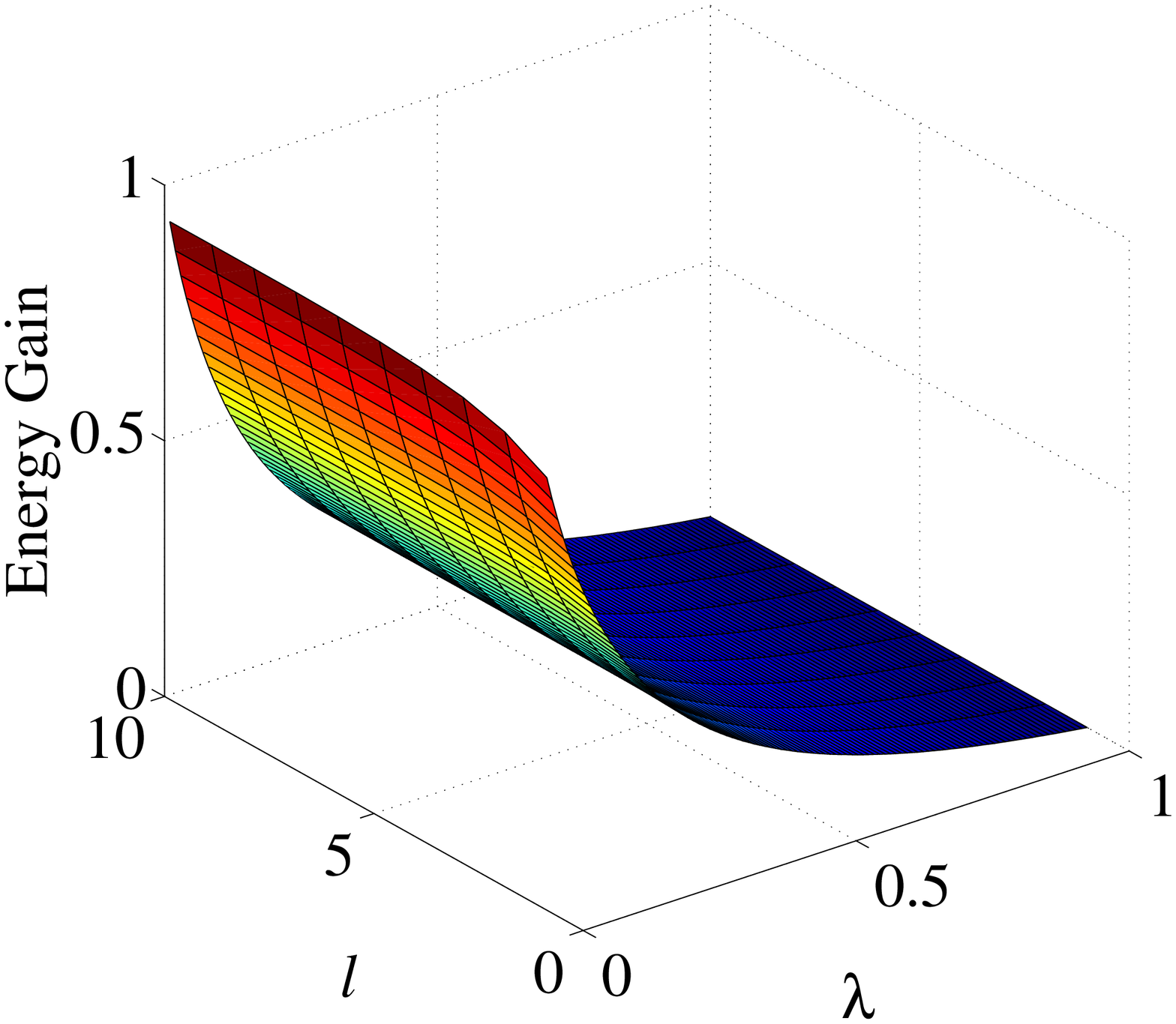,width=9pc}}
\caption{Impact of $l$ on $T$ and $G$ with either deterministic or exponential $\{S_i\}_i$.
\label{f:l-I}}
\end{center}
\end{figure*}

The third parameter used in type I like power saving classes
(scenarios D-I and D-II) is the exponent $l$. In order to assess the impact
of the maximum sleep window size on the performance of the system, we
perform a numerical analysis in which the multiplicative factor is
$a=2$, the initial window size is $T_{\min}=2$, the vacation trigger time
$T_t=0$, and the exponent $l$ is
varied from 0 to 10. We evaluate the expected sojourn time $T$ and the
energy gain $G$ both for deterministic (scenario D-I) and exponentially
distributed (scenario E-I) sleep windows. We report the results in
Fig.~\ref{f:l-I}.
Figures~\ref{f:l-T-DI} and~\ref{f:l-G-DI} respectively depict the expected sojourn time $T$ and the energy gain $G$ against the traffic input rate $\lambda$ and the exponent $l$ when sleep windows are
deterministic. The results obtained when the sleep windows are exponentially distributed are displayed in Figs.~\ref{f:l-T-EI} and~\ref{f:l-G-EI}. 

Alike the multiplicative factor, the exponent $l$ has a large impact on $T$ only for a very low traffic input rate, and has no impact on $G$ whatever the rate $\lambda$. Observe in Fig.~\ref{f:l-T-DI} that $T$ becomes almost insensitive to $l$ beyond $l=7$ (for small $\lambda$). Here the initial vacation window $T_{\min}$ is 2. We have computed $T$ considering larger values of $T_{\min}$, and have observed that $T$ saturates faster with $l$ when the initial sleep window is larger. A similar behavior is observed in the exponential case for higher $T$; cf. Fig.~\ref{f:l-T-EI}.
\subsubsection{Impact of Vacation Trigger Time $T_t$}
\label{s:T_t}
\begin{figure*}[tb]
\begin{center}
\subfigure[sojourn time, deterministic sleep windows]{
\label{f:T_t-T-DI}
\epsfig{file=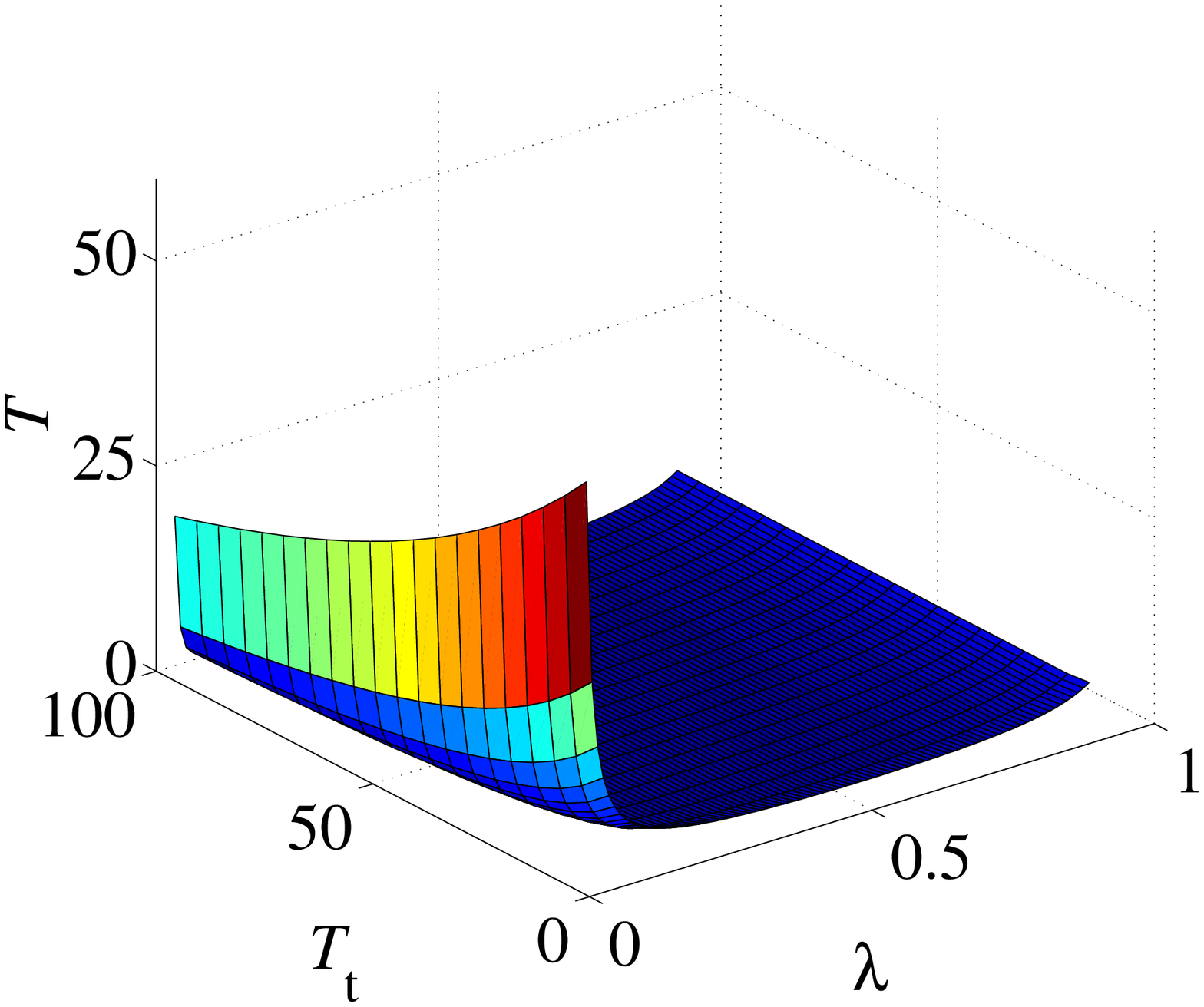,width=9pc}}
\subfigure[sojourn time, exponential sleep windows]{
\label{f:T_t-T-EI}
\epsfig{file=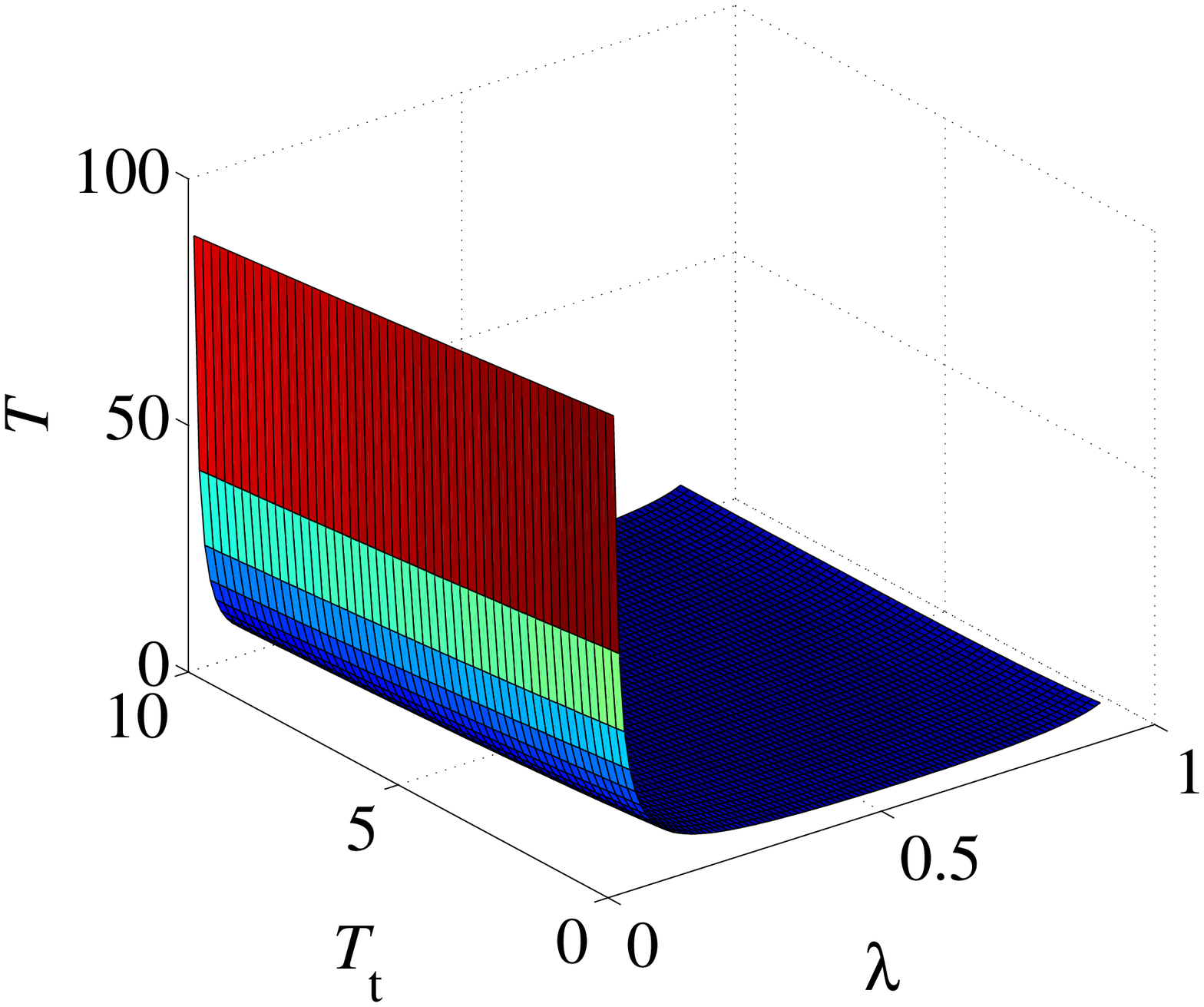,width=9pc}}
\subfigure[energy gain, deterministic sleep windows]{
\label{f:T_t-G-DI}
\epsfig{file=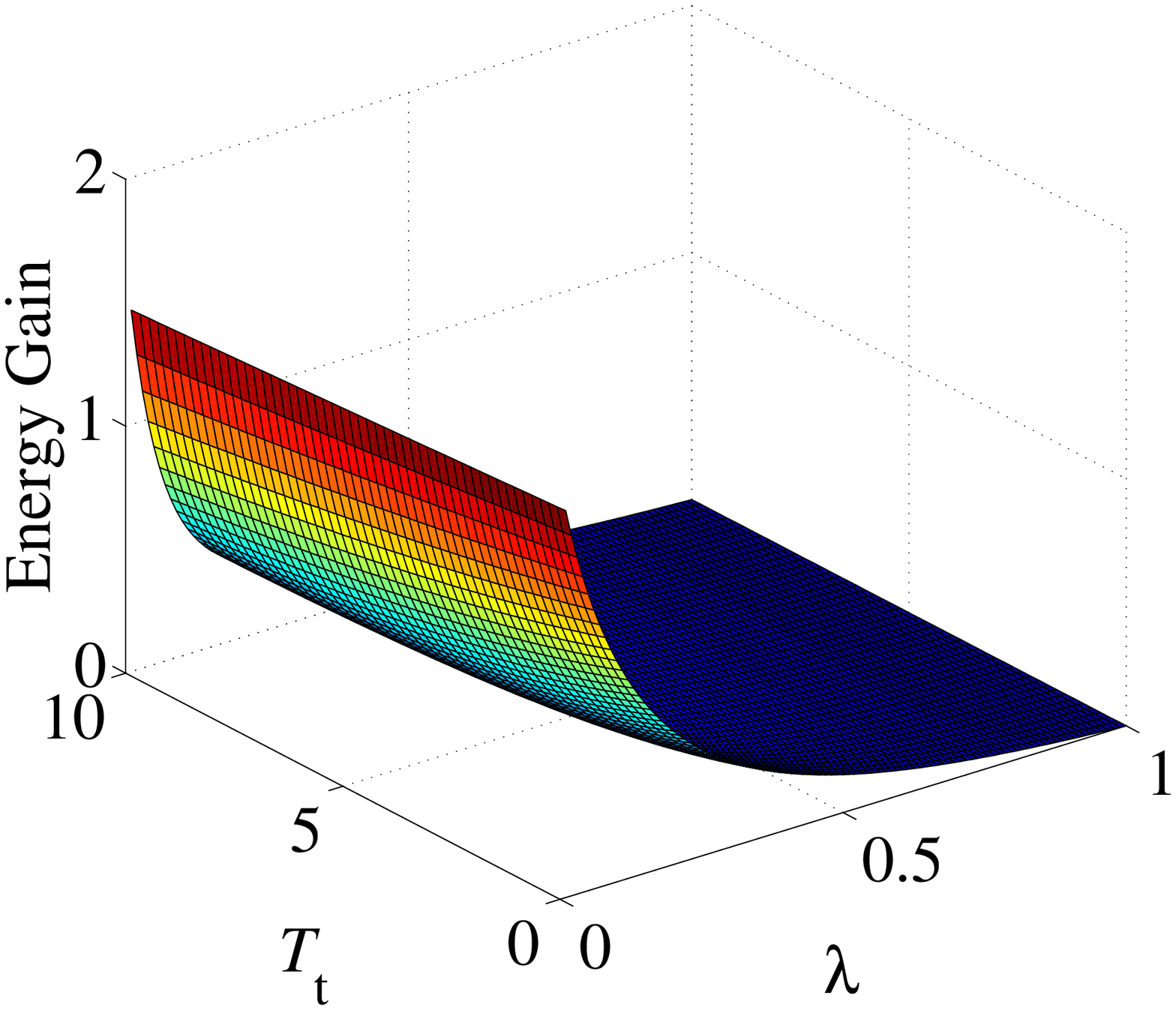,width=9pc}}
\subfigure[energy gain, exponential sleep windows]{
\label{f:T_t-G-EI}
\epsfig{file=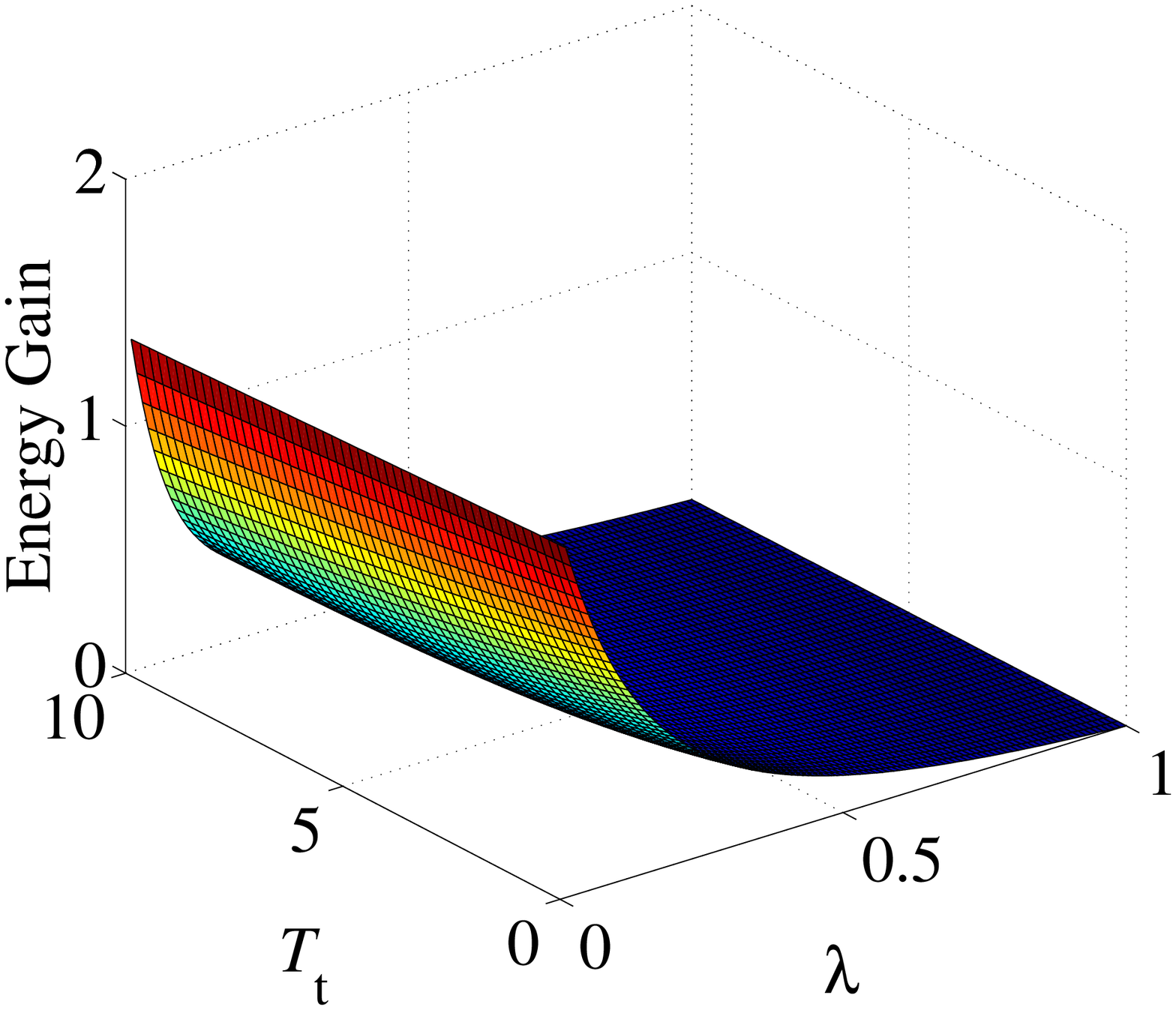,width=9pc}}
\caption{Impact of $T_t$ on $T$ and $G$ with either deterministic or exponential $\{S_i\}_i$.
\label{f:T_t-I}}
\end{center}
\end{figure*}

The fourth and the last parameter used in type I like power saving classes
is the vacation trigger time $T_t$. In order to assess the impact of the vacation trigger time on the performance of the system, we perform a numerical analysis in which the multiplicative factor is $a=2$, the initial window size is $T_{\min}=2$ the exponent $l=9$ and the vacation trigger time is varied from 0 to 10. We evaluate the expected sojourn time $T$ and the energy gain $G$ both for deterministic (scenario D-I) and exponentially distributed (scenario E-I) sleep windows. We report the results in Fig.~\ref{f:T_t-I}.
Figures~\ref{f:T_t-T-DI} and~\ref{f:T_t-G-DI} respectively depict the expected sojourn time $T$ and the energy gain $G$ against the traffic input rate $\lambda$ and the vacation trigger time $T_t$ when sleep windows are deterministic. The results obtained when the sleep windows are exponentially distributed are displayed in Figs.~\ref{f:T_t-G-EI} and~\ref{f:T_t-T-EI}.

As expected, decreasing vacation trigger time enhances the probability of the system to go on vacation resulting larger response time $T$ and larger gain $G$ as well for any $\lambda$.

\subsubsection{The expected sojourn time $T$}
\label{s:T}
The numerical results of the expected sojourn time $T$ are reported in Figs.~\ref{f:Tmin-I}--\ref{f:l-I}, parts (a) and (b). As already mentioned, $T$ is fairly insensitive to parameters $l$ and $a$ except
for very small values of $\lambda$. However, $T$ increases linearly as $T_{\min}$ increases. In scenarios D-I, E-I and E-II, as $\lambda$ increases, $T$ first decreases rapidly then becomes fairly insensitive to $\lambda$ up to a certain point beyond which $T$ increases abruptly. This can easily be explained. The sojourn time is essentially composed of two main components: the delay incurred by the vacations of the server and the queueing delay once the server is active. As the input rate increases, the first component decreases while the second one increases. For moderate values of $\lambda$, both components balance each other yielding a fairly insensitive sojourn time. The large value of $T$ at small $\lambda$ is mainly due to the ratio $\E[I_a]/\E[I]$ (recall~\eqref{e:T-alt}), whereas the abrupt increase in $T$ at large $\lambda$ is due to the term $\frac{\lambda\E[\sigma^2]}{2(1-\rho)}$, which is the waiting time in the $M/G/1$ queue without vacation.

The situation in scenario D-II is different in that $T$ is not large at small input rates $\lambda$. Recall that in this scenario, all sleep window are equal to a constant $T_{\min}$. As a consequence, the delay incurred by the vacations of the server is not as large as in the other scenarios. The balance between the two main components of the sojourn time stretches down to small values of $\lambda$.
\subsubsection{The expected energy gain $G$}
\label{s:G}
The numerical results of the expected energy gain $G$ are reported in Figs.~\ref{f:Tmin-I}--\ref{f:l-I}, parts (c) and (d). As already mentioned, $G$ is insensitive to parameters $l$ and $a$ for any $\lambda$, and sensitive to $T_{\min}$ up to a certain initial sleep window size. 

The expected energy gain $G$ decreases monotonically as $\lambda$ increases which can be explained as follows. The larger the input traffic rate $\lambda$, the shorter we expect the idle time to be and hence the smaller the gain.
\subsection{Constrained Optimization Problem}
\label{s:ResOptim}
\begin{figure}[tbh]
\begin{center}
\subfigure[scenario D-I, $T_{\textrm{QoS}}=50$]{
\label{f:optima}
\includegraphics[angle=270,scale=0.99]{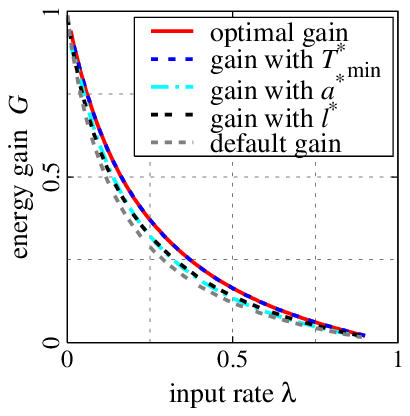}}
\hspace{1em}
\subfigure[scenario D-II, $T_{\textrm{QoS}}=50$]{
\label{f:optimb}
\includegraphics[angle=270,scale=0.99]{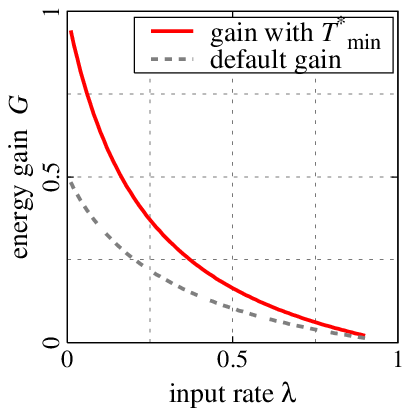}}
\hspace{1em}
\subfigure[scenario E-I, $T_{\textrm{QoS}}=100$]{
\label{f:optimc}
\includegraphics[angle=270,scale=0.99]{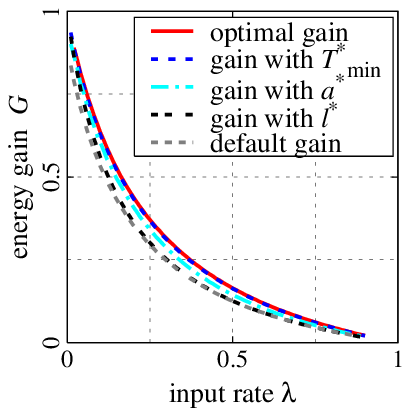}}
\hspace{1em}
\subfigure[scenario E-II, $T_{\textrm{QoS}}=100$]{
\label{f:optimd}
\includegraphics[angle=270,scale=0.99]{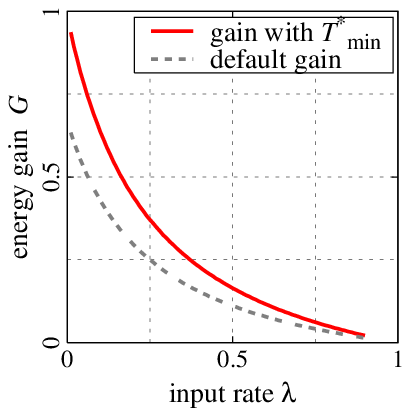}}
\caption{Maximized/default gain versus the input rate
$\lambda$.\label{f:optim}}
\end{center}
\end{figure}

\begin{figure}[tb]
\begin{center}
\subfigure[optimal $a^*$ and $l^*$ versus $\lambda$]{
\label{f:opt-a-l}
\includegraphics[angle=270,scale=0.99]{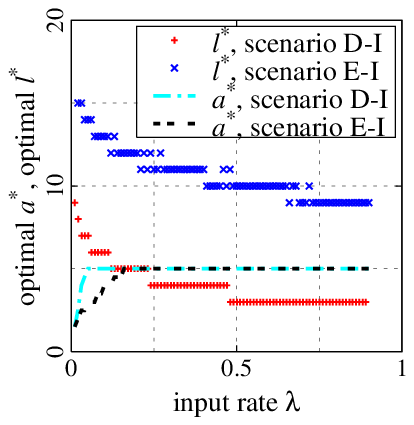}}
\hspace{1em}
\subfigure[optimal $T_{\min}^*$ versus $\lambda$ ]{
\label{f:optTmin}
\includegraphics[angle=270,scale=0.99]{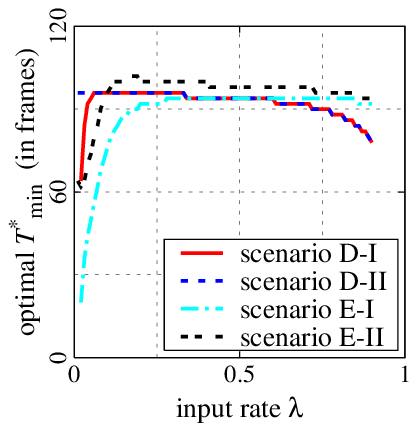}}
\hspace{1em}

\subfigure[deterministic sleep windows]{
\label{f:Ezeta-D}
\includegraphics[angle=270,scale=0.99]{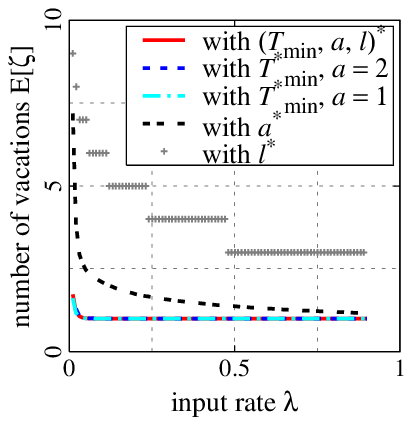}}
\hspace{1em}
\subfigure[exponential sleep windows]{
\label{f:Ezeta-E}
\includegraphics[angle=270,scale=0.99]{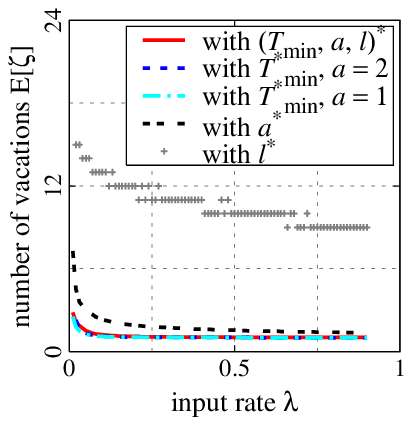}}
\caption{ Optimal parameters obtained from ${\cal P}_1$--${\cal P}_3$ and set to obtain $\E[\zeta]$ versus $\lambda$. \label{f:Ezeta}\label{f:valuesOptim}}

\end{center}
\end{figure}

We have solved the constrained optimization program introduced in Sect.~\ref{s:optim} as follows
\begin{itemize}
\item ${\cal P}_1$ for $T_{\min}^*$ when $a=2$ and $l=9$ (default
values) with $T_{\textrm{QoS}}=50$ for scenario D-I and
$T_{\textrm{QoS}}=100$ for scenario E-I, and when $a=1$ or $l=0$ with
$T_{\textrm{QoS}}=50$ for scenario D-II and $T_{\textrm{QoS}}=100$ for
scenario E-II;
\item ${\cal P}_2$ for $a^*$ with $T_{\min}=2$ and $l=9$ (default
values) with $T_{\textrm{QoS}}=50$ for scenario D-I and
$T_{\textrm{QoS}}=100$ for scenario E-I;
\item ${\cal P}_3$ for $l^*$ when $T_{\min}=2$ and $a=2$ (default
values) with $T_{\textrm{QoS}}=50$ for scenario D-I and
$T_{\textrm{QoS}}=100$ for scenario E-I;
\item ${\cal P}_4$ for $(T_{\min},a,l)^*$ with $T_{\textrm{QoS}}=50$
for deterministic sleep windows and $T_{\textrm{QoS}}=100$ for exponential sleep windows.
\end{itemize}
The optimal gain achieved by the four programs ${\cal P}_1$--${\cal P}_4$ and the gain obtained when using the default values are illustrated in Fig.~\ref{f:optim} against the input rate $\lambda$, for deterministic (Figs.~\ref{f:optima} and~\ref{f:optimb}) and exponential (Figs.~\ref{f:optimc} and~\ref{f:optimd}) sleep windows. The right-hand-side graphs depict the optimal gain (returned by program ${\cal P}_1$ when $a=1$) and the gain achieved under the default protocol parameter ($T_{\min}=2$). 
The most relevant observation to be made on each of Figs.~\ref{f:optima} and~\ref{f:optimc} is the match between the curve labeled ``optimal gain'' (result of program ${\cal P}_4$) and the curve labeled ``gain with $T_{\min}^*$'' (result of program ${\cal P}_1$). The interest of this observation comes from the fact that ${\cal P}_4$ involves a multivariate optimization whereas ${\cal P}_1$ is a much simpler single variate program.
The explanation for this match is as follows. The program ${\cal P}_1$ is being solved for the optimal $T_{\min}$. It thus quickly reduces the number of vacations $\E[\zeta]$ to $1$ (refer to Fig.~\ref{f:Ezeta}) and thereby makes the role of both $a$ and $l$ insignificant. Hence, the energy gain maximized by ${\cal P}_1$ tends to the optimal gain returned by ${\cal P}_4$.

Comparing the optimal values of $T_{\min}$ as returned by programs ${\cal P}_1$ and ${\cal P}_4$ in the deterministic case (cf. Fig. \ref{f:Ezeta} When maximizing the gain by optimizing $T_{\min}$ (program ${\cal P}_1$; see Fig.~\ref{f:optTmin}), we observe in all scenarios but scenario D-II that, optimally, $T_{\min}$ should first increase with the input rate $\lambda$ then decrease with increasing $\lambda$ for large values of $\lambda$. This observation is rather counter-intuitive and we do not have an explanation for it at the moment. Our intuition that $T_{\min}$ should decrease as $\lambda$ increases is confirmed only in scenario D-II.

Looking at the expected number of vacations $\E[\zeta]$, should the optimal value $T_{\min}^*$ be used,  it appears that $\E[\zeta]$ decreases asymptotically to 1 as $\lambda$ increases; see Fig.~\ref{f:Ezeta}. The reason behind this is the energy consumption during listen windows and warm-up periods. To maximize the energy gain, one could minimize the factor multiplying $C_{\textrm{sleep}}$, in other words minimize 
$\E[\zeta]$. As a consequence, if $T_{\min}$ is optimally selected, then the initial sleep window will be set large enough so that the server will rarely go for a second vacation period, thereby eliminating the unnecessary energy consumption incurred by potential subsequent listen windows. As a consequence, the multiplicative factor $a$ and the exponent $l$ will have a negligible effect on the performance of the system.
\subsection{Expectation and Worst Case Analysis}
\label{s:WorstExp}
In this section, we report the results of an expectation and a worst case analysis, considering the expected energy gain as performance metric. We will solve the problems stated in \eqref{e:EH-G},
\eqref{e:ES-G}, \eqref{e:WH-G} and \eqref{e:WS-G}. The decision variable is the initial sleep window size $T_{\min}$. Each problem is solved for each of the four scenarios defined in Sects.~\ref{s:S-const} and~\ref{s:S-exp}. We consider $a=2$ and $l=9$ in scenarios D-I and E-I. Recall that we necessarily have $a=1$ and $l=0$ in scenarios D-II and E-II. We consider that $\lambda$ may take five different values. These values and the corresponding probabilities $p(\lambda)$ are given in Table~\ref{t:lambda}. The values of the parameter $T_{\min}$ found for each of the problems are reported in Table~\ref{t:Tmin}.
\begin{table}[tbh]
\begin{center}
\caption{Distribution of the input rate $\lambda$}
\label{t:lambda}{ \scriptsize

\begin{tabular}{|l||lllll|}
\hline
$\lambda$ & 0.02 & 0.05 & 0.1 & 0.2 & 0.5 \\
\hline
$p(\lambda)$ & 0.3125 & 0.3125 & 0.1875 & 0.1250 & 0.0625 \\
\hline
\end{tabular} }
\end{center}
\end{table}
\begin{table}[tbh]
\begin{center}
\caption{Expectation/worst-case analysis: value of $T_{\min}$ (in number of frames)}
\label{t:Tmin}{ \scriptsize
\begin{tabular}{|l||c|c|c|c|}
\hline
& \multicolumn{2}{c|}{Expectation analysis} &\multicolumn{2}{c|}{Worst-case analysis} \\
Scenario & hard constraint & soft constraint & hard constraint & soft constraint \\
\hline
D-I & 65 & 92 & 64 & 64 \\
D-II & 96 & 97 & 94 & 94 \\
E-I & 22 & 50 & 21 & 21 \\
E-II & 69 & 79 & 62 & 62 \\
\hline
\end{tabular} }
\end{center}
\end{table}
As expected, the soft constraint allows larger value of $T_{min}$ as compared to the case of hard constraint. Since the vacation periods are of same duration for type II (D-II and E-II) policies, the difference between soft constraint to hard constraint cases are squeezed. We note that the problem of "Expected analysis" minimizes the average of sleep gain while the "worst case analysis" solves the Maxmin of the Sleep gain. Obviously, the later case will have less variability which is reflected by the almost no difference of $T_{min}$ for their hard constraint and soft constraint case results. 
\section{Conclusion and Perspectives}
\label{s:conc}
In this paper, we have analyzed the $M/G/1$ queue with repeated inhomogeneous vacations. In all prior work, repeated vacations are assumed to be i.i.d., whereas in our model the duration of a repeated vacation can come from an entirely different distribution. Using transform-based analysis, we have derived various performance measures of interest such as the expected system response time and the gain from idling the server. We have applied the model to study the problem of power saving for mobile devices. The impact of the power saving strategy on the network performance is easily studied using our analysis. We have formulated various constrained optimization problems aimed at determining optimal parameter settings. We have performed an extensive numerical analysis to illustrate our results, considering four different strategies of power saving having either deterministic or exponentially distributed sleep durations. We have found that the parameter that most impacts the performance is the initial sleep window size. Hence, optimizing this parameter solely is enough to achieve quasi-optimal energy gain.

In this paper, we have focused on deriving the expected sojourn time. However, it is possible to derive stronger results in means of the distribution of the sojourn time using the decomposition properties obtained in ~\cite{Fuhrmann_Copper_1985} and the distributional form of Little's law~\cite{KS83}. The queue length decomposition property ~\cite{Fuhrmann_Copper_1985} states that the queue length in an $M/G/1$ queue with vacations at an arbitrary epoch (i.e. in stationary regime) is distributed as the independent sum of (i) the queue length in the corresponding $M/G/1$ queue without vacation and (ii) the queue length in the $M/G/1$ queue with vacations at an arbitrary epoch during a non-busy interval. Given that our vacations are inhomogeneous, a significant portion of the derivations shall need to be repeated. However, we think it is worthy to investigate this approach and plan to do so in the near future. 

Other important research directions are considered. Namely,
\paragraph{Other traffic profiles}
It is interesting to consider more bursty real time traffic as well as TCP traffic. We expect that much of this work may have to be performed through simulations as the queueing analysis may become intractable. It is important to examine how our optimized parameters perform when a new type of traffic is introduced, and whether our robust design for the worst case Poisson traffic maintains its robustness beyond the Poisson arrival processes.
\paragraph{Extensions of the protocol}
So far our analysis enabled us to optimize parameters of the protocol. It is of interest to go beyond the optimization and to examine extensions or improvements of the protocol that would require to extend the theoretical framework as well. In particular we intend to examine rendering $T_{\min }$ dynamic, by choosing its value at the $n$th idle time as a function of the $V_\zeta$ (or of its expectation) in
the $(n-1)$-th idle time.

\bibliographystyle{IEEEtran}
\bibliography{ref_dec}

\appendix
\subsection{ Computation of Initial queue length distribution}
Denoting $(1-{\cal L}_{T_t}(\lambda))$ by ${\cal L}^c_{T_t}(\lambda))$, the z transform of initial queue size $Z(.)$ is given using eq. \eqref{e:ZI_z} by
\bear
\nonumber &&\hspace{-8mm}N_I(z)\\
\nonumber&=&\sum_{m=0}^{\infty}z^m \P(N_I=m)=z\P(N_I=1)+\sum_{m=2}^\infty z^m \P(N_I=m) \\
\nonumber &=&z {\cal L}^c_{T_t}(\lambda)+\sum_{m=1}^\infty z^m\sum_{i=1}^{\infty} \E\Big[\exp(-\lambda V_i)\frac{(\lambda V_i)^m}{m!}\Big]\\
&&\hspace{40mm}\nonumber{\cal L}_{\widehat{i-1}}(\lambda)L_{T_t}(\lambda) \\
\nonumber &=&z {\cal L}^c_{T_t}(\lambda)+\sum_{m=0}^\infty z^m\sum_{i=1}^{\infty} \E\Big[\exp(-\lambda V_i)\frac{(\lambda V_i)^m}{m!}\Big]\\
&&\hspace{40mm}\nonumber{\cal L}_{\widehat{i-1}}(\lambda)L_{T_t}(\lambda) \\
\nonumber&=&{\cal L}^c_{T_t}(\lambda) z+\sum_{i=1}^{\infty} \sum_{m=0}^{\infty}z^m\E\Big[\exp(-\lambda V_i)\frac{(\lambda V_i)^m}{m!}\Big]\\
&&\hspace{40mm}\nonumber{\cal L}_{\widehat{i-1}}(\lambda){\cal L}_{T_t}(\lambda)\\
\nonumber&=&{\cal L}^c_{T_t}(\lambda) z+\sum_{i=1}^{\infty} \E\Big[ \exp(-\lambda V_i) \sum_{m=0}^{\infty}\frac{(\lambda V_i z)^m}{m!}\Big]\\
&&\hspace{40mm}\nonumber{\cal L}_{\widehat{i-1}}(\lambda) {\cal L}_{T_t}(\lambda)\\
\nonumber&=&{\cal L}^c_{T_t}(\lambda) z+\sum_{i=1}^{\infty} \E[ \exp(-\lambda V_i) \exp(\lambda z V_i)]{\cal L}_{\widehat{i-1}}(\lambda){\cal L}_{T_t}(\lambda)\\
&=&{\cal L}^c_{T_t}(\lambda) z+\sum_{i=1}^{\infty} {\cal
L}_{i}(\lambda(1-z)) {\cal
L}_{\widehat{i-1}}(\lambda){\cal L}_{T_t}(\lambda). \eear
Since the arrival
is a Poisson process, the pgf of arrival during the fixed warm up
period $T_w$ is given as
\bear \hspace{-3mm}N_{T_{\tilde{w}}}(z)&=& \sum_{i=0}^{\infty} z^i\P(N_{T_w}=i)\\
\nonumber&&=\sum_{i=0}^\infty[\P(t_f>T_t)z^i\P(N_{T_{\tilde{w}}}=i)\\
&&+
\P(t_f\leq
T_t)z^i\P(N_{T_{\tilde{w}}}=i)]\\
N_{T_{\tilde{w}}}(z)&=& {\cal L}_{T_t}(\lambda){\cal
L}_{T_w}(\lambda(1-z))+ {\cal L}^c_{T_t}(\lambda)\eear
Where the Laplace transform of the arrivals during the warm up period $T_w$ is given is
\bear \label{e:ZW}
N_{T_{{w}}}(z)=
e^{-\lambda T_w(1-z)}:={\cal L}_{T_w}(\lambda(1-z)). \eear
Combining
both the above we can express the pgf of $Z$
\bear \label{e:Zzf}
N(z)&=&N_I(z)N_{T_{\tilde{w}}}(z)\nonumber\\
\nonumber& =&\left(z {\cal L}^c_{T_t}(\lambda)+\sum_{i=1}^{\infty}
{\cal L}_{i}(\lambda(1-z)) {\cal
L}_{\widehat{i-1}}(\lambda){\cal L}_{T_t}(\lambda)\right)~\\
&&[{\cal L}_{T_t}(\lambda){\cal
L}_{T_w}(\lambda(1-z))+ {\cal L}^c_{T_t}(\lambda)]. \eear
%
Note that $T_t=0$ correspond to have always vacation while $T_t=\infty$ forces no vacation. At $T_t=0$, we have ${\cal L}_{T_t}(\lambda)=e^{-\lambda T_t}=1$. We thus obtain $N(z)|_{T_t=0}=[\sum_{i=1^\infty} {\cal L}_{i}(\lambda(1-z)) {\cal L}_{\widehat{i-1}}(\lambda)][{\cal L}_{T_w}(\lambda(1-z))]$. This is in congruence with the earlier result obtained in \cite{questa}. However at $T_t=\infty$, we have
${\cal L}_{T_t}(\lambda)=e^{-\lambda T_t}=0$. We thus obtain $N(z)|_{T_t=\infty}=[z]$, which is again true as there is only one arrival as in standard M/G/1 queue.
\subsection{Computation of Sojourn time}
\label{sec:a_sojourn_time}
We assume (general assumption) that the waiting time of a customer is independent of the part of the arrival process that occurs after the customer's arrival epoch. Our policy, which is FCFS discipline, falls in this category.  The waiting time of an arbitrary customer in a queue is exactly the number of customer ahead of the tagged customer in the queue under FCFS scheme. The number of customers left behind by the tagged customers is precisely the number of arrivals during the sojourn time (waiting + service) of the tagged customers, denoted its pgf by $N(z)$. Since Poisson arrival see time average (PASTA, see Wolf), the pgf of number customer ahead of a random customer has the same pgf as $N(z)$. Therefore we can express the LST of the waiting time $W^*(s)$ of a random customer in the queue as (from \cite{Fuhrmann_Copper_1985}) 
\bear
W^*(s)&=& \frac{\lambda[1-N(1-s/\lambda)]}{s \E[N]} W^*_{M/G/1}(s).
\eear
Where, $W^*_{M/G/1}(s)$ is the LST of waiting time of an arbitrary request in the queue (excluding its service time) of a standard $M/G/1$ queue. From \cite{Takagi}(1.45), we have $W^*_{M/G/1}(s)=\frac{s(1-\rho)}{s-\lambda+\lambda \sigma^*(s)}$, where LST of service time is given by $\sigma^*(s)=\E[e^{-s\sigma}]$. 
Thus, we have
\bear
\label{e:Wsf}
\nonumber W^*(s)&=& \frac{\lambda[1-N(1-s/\lambda)]}{s \E[N]} \frac{s(1-\rho)}{s-\lambda+\lambda \sigma^*(s)}\\
&=& K \frac{[1-N(1-s/\lambda)]}{s-\lambda+\lambda \sigma^*(s)}.
\eear
where $K=\frac{(1-\rho)\lambda}{\E[N]}$.
The moments of the $W(.)$ can be obtained from its LST by simply evaluating its derivatives at $s=0$, i.e., $E[W^{*n}]=(-1)^nW^{*{(n)}}(0)$.
The Expected waiting time is the first moment, given by $\E[W]=-W^{*(1)}(0)$. This can be computed from \eqref{e:Wsf}
by a routine but tedious calculation (two applications of
L'Hospital's rule is required),
\bear
\nonumber &&\hspace{-7mm}{W}^{*(1)}(s)(s-\lambda+\lambda \sigma(s))+W^*(s)(1+\lambda {\sigma^{(1)}}(s))\\
\nonumber&&\hspace{7mm}=K/\lambda [{N^{(1)}}(1-s/\lambda)],\\
\nonumber&&\hspace{-7mm}{W}^{*(2)}(s)(s-\lambda+\lambda \sigma(s))+2{W}^{*(1)}(s)(1+\lambda {\sigma^{(1)}}(s))\nonumber\\
&&\hspace{25mm}+W^{*}(s)\lambda {\sigma^{(2)}}(s)\nonumber\\
&&\hspace{7mm}=-K/\lambda^2 [{N^{(2)}}(1-s/\lambda)].
\eear
Evaluating the above at $s=0$ and using the following: i) $W^*(0)=\sigma(0)=1$ (from the definition of LST); ii) $1+\lambda {\sigma^{(1)}}(0)=1-\rho$; and, iii) $K=\frac{(1-\rho)\lambda}{\E[N]}$; we obtain
\bear
\nonumber&&2{W}^{*(1)}(0)(1-\rho)+\lambda {\sigma^{(2)}}(0)=-\frac{(1-\rho)}{\lambda\E[N]}\left({N^{(2)}}(1)\right),\\
&&\E[W]=-{W}^{*(1)}(0)=\frac{{N^{(2)}}(1)}{2\lambda\E[N]}+
\frac{\lambda\E[\sigma^2]}{2(1-\rho)}~.
\eear
substituting $N^{(2)}(1)$ from eq. \eqref{e:Z''1} and $\E[N]$ eq.
\eqref{e:EZ} we obtain the mean waiting time as
\bear
\E[W]=\frac{T_w^2{\cal L}_{T_t}(\lambda)+2T_w{\cal L}_{T_t}(\lambda)\E[\tilde{I}]+\E[I_a]}{2(T_w{\cal L}_{T_t}(\lambda)+\E[\tilde{I}])}
+\frac{\lambda \E[\sigma^2]}{2(1-\rho)}.
\eear
For the second moment, we perform the derivative one more time,
\bear
&&\hspace{-7mm}\nonumber{W^{*(3)} }(s)(s-\lambda+\lambda \sigma(s))+3{W^{*(2)}}(s)(1+\lambda {\sigma^{(1)}}(s))\\
&&\hspace{-5mm}+3{W^{*(1)}}(s)\lambda {\sigma^{(2)}}(s) +W^*(s)\lambda {\sigma^{(3)}}(s)=K/{\lambda^3} [{N^{(3)}}(1-s/\lambda)]. \hspace{5mm}
\eear
Evaluating at $s=0$ we obtain the second moment,
\bear
\E[W^2]=W^{*(2)}(0)=\frac{N^{(3)}(1)}{3\lambda^2\E[N]}+\frac{\lambda \E[W] \E[\sigma^{2}]}{(1-\rho)} +\frac{\lambda \E[\sigma^{3}]}{3(1-\rho)}
\eear
Substituting $N^{(3)}(1)$ from eq. \eqref{e:Z'''1f}, finally we have
\bear
\nonumber\E[W^2]&=&\frac{ T_w^3 {\cal L}_{T_t}(\lambda)+\E[I_c]+3 (T_w^2{\cal L}_{T_t}(\lambda)\E[\widetilde{I}] +{\cal L}_{T_t}(\lambda) \E[I_a])}{3(T_W{\cal L}_{T_t}(\lambda)+\E[\widetilde{I}])}\\
&& +
\frac{\lambda \E[\sigma^{2}]}{(1-\rho)}\E[W]+\frac{\lambda \E[\sigma^{3}]}{3(1-\rho)}
\eear
\subsection{Third moment of Initial Queue distribution and sojourn time}
Moving further, the third derivative of the $N(.)$ can be obtained similarly from eq.
\eqref{e:Z''z},
\bear
\label{e:Z'''z}
N^{(3)}(z)= N_I(z)N^{(3)}_{T_{\tilde{w}}}(z) + N^{(3)}_I(z)N_{T_{\tilde{w}}}(z) + 3[ N^{(1)}_I(z) N^{(2)}_{T_{\tilde{w}}}(z)&&\nonumber\\
+N^{(2)}_I(z) N^{1}_{T_{\tilde{w}}}(z)] .&&\\
N^{(3)}(1)= N_I(1)N^{(1)}_{T_{\tilde{w}}}(1) + N^{(3)}_I(1)N_{T_{\tilde{w}}}(1) + 3[ N^{(1)}_I(1) N^{(2)}_{T_{\tilde{w}}}(1)&&\nonumber\\+N^{(2)}_I(1) N^{1}_{T_{\tilde{w}}}(1)] .&&
\eear
Using eq. \eqref{e:Z''z} we can obtain
\bear
\label{e:Z'''z1}
N^{(3)}_I(z)&=&\sum_{i=1}^{\infty} \E[(\lambda V_i)^3]{\cal L}_{i}(\lambda(1-z)){\cal L}_{\widehat{i-1}}(\lambda){\cal L}_{T_t}(\lambda) .\\
\label{e:Z'''1}
N^{(3)}_I(1)&=&\sum_{i=1}^{\infty} \E[(\lambda V_i)^3] {\cal L}_{\widehat{i-1}}(\lambda){\cal L}_{T_t}(\lambda)=\lambda^3\E[I_c].
\eear
Where we denote $\E[I_c]:=\E[V_i^3] {\cal L}_{\widehat{i-1}}(\lambda){\cal L}_{T_t}(\lambda)$.
Using eq. \eqref{e:EIa1}, we obtain
\bear
\label{e:Z'''zw}
N^{(3)}_{T_{\tilde{w}}}(z)&=& (\lambda T_w)^3 {\cal L}_{T_t}(\lambda) {\cal L}_{T_w}(\lambda(1-z)).\\
\label{e:Z'''1w}
N^{(3)}_{T_{\tilde{w}}}(1)&=& (\lambda T_w)^3 {\cal L}_{T_t}(\lambda)
\eear
Combining from eq. \eqref{e:Z'''z1}- \eqref{e:Z'''1w}, we can express the
\bear
\label{e:Z'''1f}
&&\hspace{-8mm}N^{(3)}(1)\nonumber\\
&&\hspace{-5mm}= \lambda^3\left( T_w^3 {\cal L}_{T_t}(\lambda)+\E[I_c] +3[ T_w^2 {\cal L}_{T_t}(\lambda)\E[\tilde{I}]+\E[I_a]{\cal L}_{T_t}(\lambda)\right).\hspace{4mm}
\eear
\noindent The third moment is given by the relation $\E[N^3]=N^{(3)}(1)+3\E[N^2]-2\E[N]$.
The third moment of the sojourn time can be obtained similarly as above
\bear
\E[W^2]=\frac{N^{(3)}(1)}{3\lambda^2\E[N]}+\frac{\lambda \E[W] \E[\sigma^{2}]}{(1-\rho)} +\frac{\lambda \E[\sigma^{3}]}{3(1-\rho)} \eear
\bear
\nonumber \E[W^2]&=&\frac{ T_w^3 {\cal L}_{T_t}(\lambda)+\E[I_c]+3 (T_w^2{\cal L}_{T_t}(\lambda)\E[\widetilde{I}] +{\cal L}_{T_t}(\lambda) \E[I_a])}{3(T_W{\cal L}_{T_t}(\lambda)+\E[\widetilde{I}])}\\&&+
\frac{\lambda \E[\sigma^{2}]}{(1-\rho)}\E[W] +\frac{\lambda \E[\sigma^{3}]}{3(1-\rho)}
\eear
\end{document}